\documentclass[lettersize,journal]{IEEEtran}
\usepackage{amsmath,amsfonts}
\usepackage{algorithmic}
\usepackage{algorithm}
\usepackage{array}
\usepackage[caption=false,font=normalsize,labelfont=sf,textfont=sf]{subfig}
\usepackage{textcomp}
\usepackage{stfloats}
\usepackage{url}
\usepackage{enumerate}
\usepackage{verbatim}
\usepackage{graphicx}
\usepackage{xcolor}
\usepackage{cite}
\usepackage{tabularx,booktabs,textcomp}
\usepackage{amsthm}

\usepackage{amsmath,amssymb,amsthm}

\begin{document}

\title{Modelling of the Electric Vehicle Charging Infrastructure as Cyber Physical Power Systems: A Review on Components, Standards, Vulnerabilities and Attacks}

\author{{Sagar Babu Mitikiri, K. Victor Sam Moses Babu, Divyanshi Dwivedi, Vedantham Lakshmi Srinivas, Pratyush Chakraborty, Pradeep Kumar Yemula, Mayukha Pal$^{*}$}

\thanks{$^{*}$(Corresponding author: Mayukha Pal)}

\thanks{Mr. Sagar Babu Mitikiri is a Data Science Research Intern at ABB Ability Innovation Center, Hyderabad 500084, India, and also a Research Scholar at the Department of Electrical Engineering, Indian Institute of Technology (ISM), Dhanbad 826004, IN.}
\thanks{Mr. K. Victor Sam Moses Babu is a Data Science Research Intern at ABB Ability Innovation Center, Hyderabad 500084, India and also a Research Scholar at the Department of Electrical and Electronics Engineering, BITS Pilani Hyderabad Campus, Hyderabad 500078, IN.}
\thanks{Mrs. Divyanshi Dwivedi is a Data Science Research Intern at ABB Ability Innovation Center, Hyderabad 500084, India, and also a Research Scholar at the Department of Electrical Engineering, Indian Institute of Technology, Hyderabad 502205, IN.}
\thanks{Dr. Vedantham Lakshmi Srinivas is an Asst. Professor with the Department of Electrical Engineering, Indian Institute of Technology (ISM), Dhanbad 826004, IN.}
\thanks{Dr. Pratyush Chakraborty is an Asst. Professor with the Department of Electrical and Electronics Engineering, BITS Pilani Hyderabad Campus, Hyderabad 500078, IN.}
\thanks{Dr. Pradeep Kumar Yemula is an Assoc. Professor with the Department of Electrical Engineering, Indian Institute of Technology, Hyderabad 502205, IN.}
\thanks{Dr. Mayukha Pal is with ABB Ability Innovation Center, Hyderabad-500084, IN, working as Global R\&D Leader – Cloud \& Analytics (e-mail: mayukha.pal@in.abb.com).}
}


\maketitle
\thispagestyle{empty}
\begin{abstract} 
The increasing number of electric vehicles (EVs) has led to the growing need to establish EV charging infrastructures (EVCIs) with fast charging capabilities to reduce congestion at the EV charging stations (EVCS) and also provide alternative solutions for EV owners without residential charging facilities. The EV charging stations are broadly classified based on i) where the charging equipment is located - on-board and off-board charging stations, and ii) the type of current and power levels - AC and DC charging stations. The DC charging stations are further classified into fast and extreme fast charging stations. This article focuses mainly on several components that model the EVCI as a cyber-physical system (CPS). The various components of EVCI include energy storage systems, EV supply equipment, human-machine interfaces, communication systems, charging station management systems, etc. These components are interconnected with real-time processing units, communication networks, sensors, and actuators, thus making EVCI a CPS. Due to the data flow in the CPS, it is essential to maintain privacy, safety, integrity, and correctness of parameters at each stage. Several standards and protocols are defined to protect the EVCI from various threats and cyber attacks to ensure reliable and safe operation in both grid-connected and islanded modes. But still, there exist several vulnerabilities in the EVCI due to communication requirements at almost every stage. These vulnerabilities can lead intruders and adversaries to infiltrate the system and potentially cause damage through cyber attacks. Smart and digital payment schemes also introduce additional security concerns. Consequently, it is crucial to employ effective attack detection and impact mitigation methods. Thus, this paper examines the various standards and protocols governing communication establishment and controlled regulated power distribution. Furthermore, the defense and detection methods for various vulnerabilities and attacks in the EVCI are comprehensively presented, offering robust measures to secure this vital infrastructure. 
\end{abstract}

\begin{IEEEkeywords}
CPPS, EVCI, Fast Charging, CHAdeMO, FDIA, Communication Protocols, Cyber attacks, Detection, Mitigation, Prevention.
\end{IEEEkeywords}

\section{Introduction}
\label{section:Introduction}

In general, a CPS is an intelligent system comprising both a physical unit and a cyber unit. By combining interaction, processing, control, and computing, it integrates physical and computational capabilities. The physical unit of CPS comprises of various data-collecting elements and sensors that communicate with the cyber unit of the system. The cyber unit processes this data, analyzes statistical information, performs computations, and generates an operating signal that is transmitted back to the physical system via the communication network \cite{1}. The perception layer, the transport layer, and the application layer are the three layers that make up CPS. The perception layer is utilized to gather insightful data and put feedback decisions into action. The cyber unit, medium access control (a medium for connecting the cyber and physical units), and physical unit are all included in the transport layer, which is utilized to transmit information and decisions made regarding the various components. The application layer, which is also known as the control layer, is used primarily to make decisions based on the outcomes of the analysis of perceptive information \cite{2}.

To significantly enhance the capabilities of physical systems, CPS utilizes computer hardware, software, and communication networks that are integrated and engaged with the ongoing tasks \cite{CROWDER2020271}. Due to the large interaction between the physical and cyber (control and communication) units of the CPS, there exists a large number of vulnerable areas which leads the adversaries to cause severe attacks on the system that lead to major consequences. The applications of CPS technology have been observed in various fields such as transportation, robotics, defense, aviation, and critical infrastructure \cite{wang2015current}.

\subsection{Real World Scenarios}
Many scenarios of attacks on the CPS exist in the real world. The expansion of applications of cyber systems in different disciplines results in an increase in the various vulnerable areas. These vulnerable areas increase the likelihood of adversaries, launching various types of cyber attacks. A few incidents and blackouts resulting from CPS attacks in various fields are illustrated in this section to show the severity of the problem.

In 1982, the manipulation of the gas pipeline control software, which was developed by the Central Intelligence Agency (CIA), resulted in exceeded pressure limits that led to a massive explosion \cite{reed2005abyss}. In 2003, the penetration of the slammer worm into the David-Besse nuclear plant through the contractor's network resulted in the disabling of system indicators and safety parameters for 5 hours \cite{hardy2012software}. Due to this, the control operators are unable to monitor the temperature of the reactor core which is a crucial parameter. Another incident occurred where a hacker penetrated the testing generator of USA in 2007 causing a rapid succession of turn on and off of its circuit breaker that made the generator explode resulting in a loss of  1 million dollars \cite{swearingen2013you}. 

On December 23, 2015, a major blackout took place in Kyiv, Ukraine, affecting three major distribution companies and 225,000 customers for several hours. The investigations revealed that the blackout was caused due to the involvement of a cyber attack affecting seven 110KV and 23KV substations. The attackers invaded the supervisory control and data acquisition system (SCADA) and opened all the circuit breakers leaving the operator with no access for possible restoration\cite{case2016analysis}. Another cyber attack took place in Kyiv by shutting down the 200MW generation which is equivalent to 20\% of the nighttime electrical energy consumption of that place \cite{condliffe2016ukraine}. The impacts of a few other cyber are provided in Table \ref{attacks table}. Furthermore, between 2010 and 2014, the US Department of Energy (DoE) documented 150 successful assaults on systems that housed data about electrical infrastructure \cite{five}.

\begin{table*}
\centering
\caption{Impacts of cyber attacks in various countries}
\label{attacks table}
\centering
\renewcommand{\arraystretch}{1.5}
\begin{tabular}
  {p{3cm}p{7cm}p{6cm}}
    \toprule
\textbf{Year \& Location}\vspace{0.1cm}& \textbf{Descriptiion of the attack} \vspace{0.1cm}& \textbf{Impact of the attack}\vspace{0.1cm} \\
    \midrule
     1999, Bellingham \cite{Bellingham} & Slowdown of SCADA systems of a Gasoline pipeline. &  Kilotons of TNT equivalent explosion. \\ 
     2008, Turkey \cite{OSTI} & Manipulation of control system parameters of oil pipeline. & Oil explosion, 30k barrels spilled in water.\\
     2012, Saudi Arabia \& Qatar \cite{saudi} & Malware affected Aramco and Ras gas. & Generation and delivery of energy have affected.\\
     2020, US \cite{tran2021solarwinds} & Solarsinds cyber attack: Orion IT monitoring and management software. & More than 30,000 public and private organizations are affected.\\
     2022, Ukraine \cite{burke2022attacks} & Attack on nuclear power plant website Energoatom by Russian hacktivists. & Disruption of online services lasted for a few hours.\\
      2014, United States, France, Spain, Germany, Italy, Poland, and Turkey \cite{khan2023dragonfly} & Over 1000 organizations had been targeted with 84\% being energy sector. & The active cyber espionage outfit Energetic Bear has been operating with the intention of disrupting and conducting surveillance.\\
      \bottomrule
\end{tabular}
\end{table*}

Cyberattacks have increased in recent years, such as the attack on  U.S. based power utilities on March 5, 2019 \cite{de2023cybersecurity}, Kudankulam nuclear power plant attack in India \cite{dilipraj2019supposed} and the man-in-the-middle attack in Nuclear Power Corporation of India Limited (NPCIL) \cite{akalp2020analysis} on October 30, 2019. Other recent victims of the cyber attacks are Schneider Electric and Siemens Energy, which were targeted by ransomware groups the MOVEit attack. MOVEit \cite{progress} is a software tool used for the managed file transfer (MFT). On May 30, 2023, Siemens confirmed that they were victims of the attack, stating that no critical data was compromised and their operations were not affected. Schneider Electric was informed on June 26, 2023, that they had been the target of a cyberattack related to MOVEit, but Schneider Electric only stated that they were looking into this assertion \cite{Securityweek}. Further, the details and analysis of these cyber attack incidents are found in \cite{3}.

\subsection{Contribution of Work}
It is found in the literature review, that there are only a few review articles that consider the various aspects of EVCI as CPS. One of the first articles to provide a detailed review of EV charging security with a focus on the device and network-level vulnerabilities that are more frequent at EV, EV charging station (CS), and grid levels is \cite{acharya2020cybersecurity}. It also examines feasible cyber attack scenarios on EVCI and its network, assessing the technical and financial risks posed to power grids, including demand-side attacks. A critical review of cyber attacks and cybersecurity on load frequency control (LFC) of the power systems is presented in \cite{mohan2020comprehensive}. The modeling of the energy cyber physical paths and review of aligning its research paths are presented in \cite{orumwense2019systematic}. Rajaa et al. of \cite{3} describe the different modeling and simulation methods of the cyber physical power systems (CPPS) with cybersecurity analysis and applications. CPPS is an integration of the physical power system with the cyber system (or unit). These systems cover the various domains of electrical power system like generation, transmission, distribution and utilization \cite{cao2020cyber, suryanarayanan2016cyber}. Apart from these, a lot of papers are available for review on CPPS and cybersecurity applications \cite{van2022cyber, agrafiotis2018taxonomy, yohanandhan2020cyber}. Integration of renewable energy sources (RES) with EV technologies is assessed in \cite{barman2023renewable}. The contribution of this article is explained as follows:

\begin{enumerate}
    \item Several reasons and components that portray the EVCI as a CPS are presented. Furthermore, it adds a comprehensive description of the constituents within the cyber and physical layers of the EVCI, and the interaction between the layers.
    \item An architectural model for the EVCI is introduced, as depicted in Fig. \ref{EVCIstructure}. This model illustrates the key components of EVCI and outlines the pathways for communication and power flow.
    \item Charging stations are categorized according to the charging methodologies, such as on-board and off-board charging methods, providing a comprehensive evaluation of the EVCI.
    \item various standards and protocols required for the EVCI, to achieve interoperability, safety, efficiency, data communication, grid integration, and regulatory compliances are explained with their latest versions and are categorized based on the defining organizations and institutions..
    \item Several nodes in the EVCI that are vulnerable to cyber attacks are identified and classified based on the variations in the attacks that are possible to occur.
    \item The importance of cybersecurity for the EVCI by indicating the effects of cyber attacks on social and economic fronts is presented. Also, the objectives, prerequisites, and challenges associated with implementing cybersecurity measures are emphasized.
    \item The attack detection and defense techniques are paired together since the majority of detection techniques have built-in defense capabilities, such as isolating the affected areas.
    \item Intrusions, anomalies, and attacks are collectively treated as a cohesive unit, due to their similar impact on the system. Several research articles are reviewed for anomaly detection, which focuses on unavailability of the abnormal data.
\end{enumerate}
\subsection{Paper Organization}
This paper is structured so that the introduction and history of the many cyberattacks that have taken place are included in the introduction section. Section \ref{Cyber Physical Power Systems} provides the definition of CPS and CPPS, and illustrates how EVCI can be considered a CPS, and also describes the various components of EVCI. The architecture of the EVCI and the developments in the EVCI and the EV industry are discussed. Towards the later end of the section, the classification of EVCI based on the type of charging system adopted and role of communications in EVCI is explained and then a brief about the battery technologies and the mechanisms adopted is presented. The various standards and protocols used in the EVCI for different purposes such as the EVCI development, charging ports, safety measures, and communication standards are presented in Section \ref{standards and protocols}. Section \ref{Cyber Security} details the cyber security of the CPS where the objectives of the CPS are mentioned along with the impacts of the cyber attacks (both technical and social). Also, the requirements for the cyber security along with the challenges in implementing cyber security are described. Section \ref{Vulnerability discussion} explains the different vulnerable points where the occurrence of cyber-attacks is frequent. Further, different types of possible attacks are also mentioned. Section \ref{Detection and defense methods} provides a brief overview of the few attack detection and defense mechanisms. Finally, Section \ref{section:Conclusion} provides the future research gaps and the conclusions.

\section{EVCI}
\label{Cyber Physical Power Systems}

\begin{figure*}
    \centering
    \includegraphics[scale=0.23]{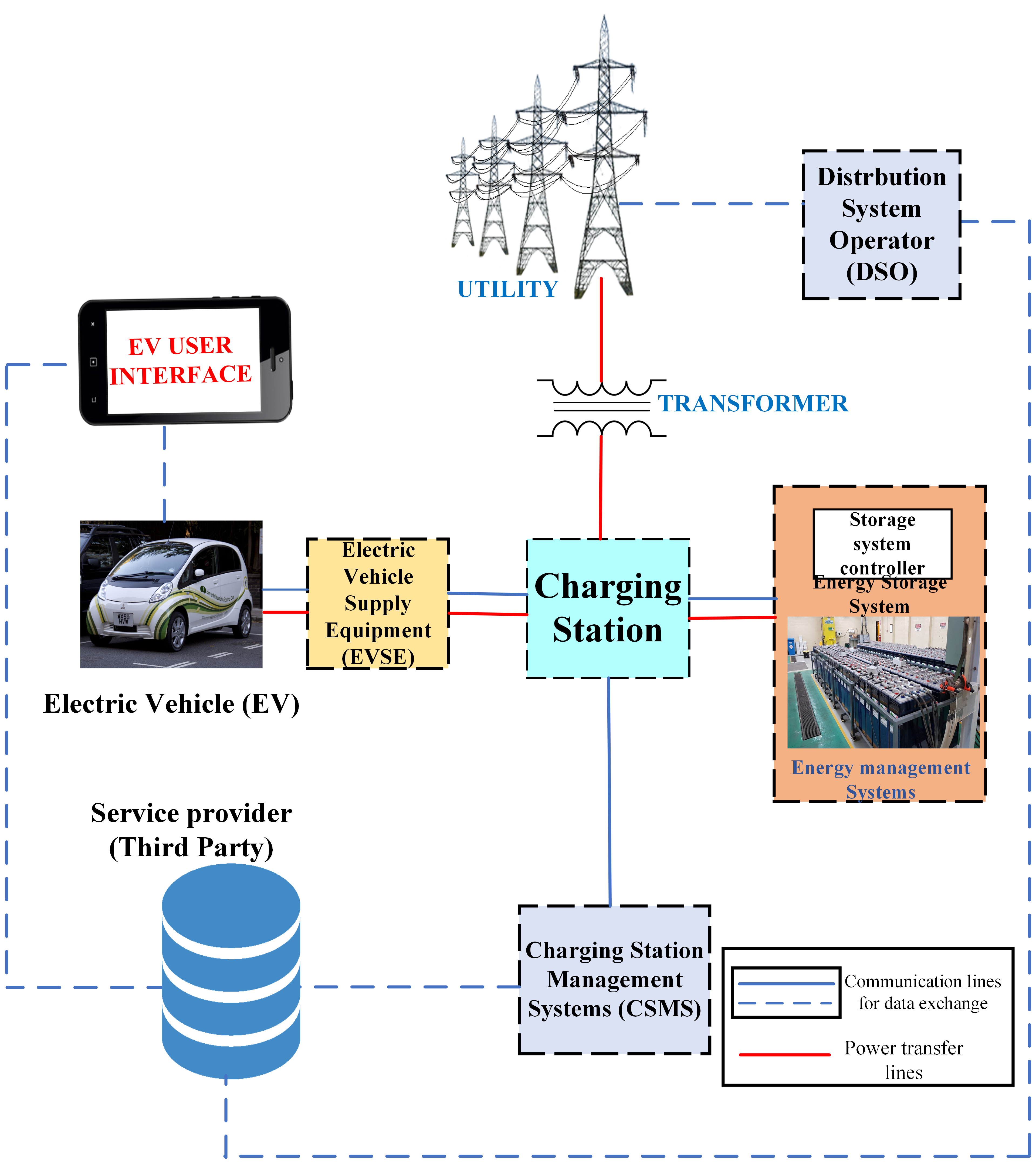}
    \caption{Structure of a typical EVCI}
    \label{EVCIstructure}
\end{figure*}

\subsection{EVCI and its Components}
The typical structural architecture of EVCI with its detailed representation is shown in Fig. \ref{EVCIstructure}. The power flow and information flow are distinguished. For more reliability and flexibility in operation, both the lines are made bi-directional in nature. The list of the architectural components of the EVCI are as follows:
\begin{enumerate}
    \item \textit{Electric Vehicle Supply Equipment (EVSE)} - It is a core sub-component of the charging station that provides the connectivity for the EV to charge and discharge \cite{6662361}. 
    \item \textit{Electric Vehicle (EV)} - TThough EVs are not always connected to the EVCI, they are considered as a critical component due to the different types of batteries available in the EVs arriving at the EVCI. Table \ref{evclassificaton} shows the different EV models and their manufacturers. Since most of the EVs store the energy in their batteries are rechargeable, the charging process of the EV is a plug-in charging type, or battery swapping type. However, the main concern for charging the EVs is the batteries. A detailed study of the technologies and other aspects of the batteries is provided in further sections. Furthermore, the EVs are classified as hybrid electric vehicles (HEV) and pure EVs \cite{das2020electric}.
    \item \textit{Charging Station Management System (CSMS)} - The main tasks of CSMS include hosting user applications, collecting and storing the user charging data, communicating with the different parts through which the power flow occurs, defining the operational parameters based on user input data, and the status of EVs and utility, and maintaining a database for the maintenance services. To ensure the quality of the power supply and also to step the voltages to a suitable level, the charging stations are connected to the utility through the transformers as shown in Fig. \ref{EVCIstructure}. The CSMS also contains the different switching panels connected to various levels and parts of the EVCI. Modern panels are operated both manually (remote and local) and automated. Additional details on switchgear and switch panels are provided in further sections.
    \item \textit{Charging Station (CS)} - It is managed by the CSMS. It controls the power flow thereby controlling the charging process, It hosts a collection of one or more charging ports. 
    \item \textit{Controller or Optimizer} - The functions of this unit are to regulate the power flow from the different power sources (grid, DERs, and storage systems). Power regulation is performed by considering different aspects such as power sharing, pricing schemes etc., and it also controls the charging power limits in the charging stations.  
    \item \textit{Distribution System Operator (DSO)} - The DSO's primary responsibility is to make sure the end-user has access to electricity. The DSO controls the flow of electricity to the charging location and grid decongestion depending on the data feedback from the EV \cite{noel2019realizing}. Similarly, a charging point operator (CPO) is an organization or person who oversees the EV charging network. The choices of CPO's are implemented by CSMS.
    \item \textit{Energy Management System (EMS)} - It is an intermediate control system that controls the power flow from the sources. It acts as an alternative power source (such as DERs, and storage systems) from the grid. If an EMS is present in the charging system, then the required power for the charging may be supplied directly from the grid or from the DERs whenever necessary.
    \item \textit{Service Provider} - These are third-party operators, consisting of the different industries that provide the interface for the EVSE owner, DSO, and CS operator. The services provided by this unit are web-based, application-based, manual-based etc., The level of the service depends upon the subscription taken by the EVSE owner, DSO, and CS operator. They offer services such as visualization, storage of information, secured data exchange between entities, payment gateways, etc.,  
    \item \textit{EV User Interface} - It is also a third-party component that is dependent on the service provider. It displays the information it gets through a mobile app or a web-based portal that helps the EV user to monitor the charge duration, current status of the EV, pricing details, and also in making digital payments. Overall, it works as human machine interface (HMI).
\end{enumerate}

\subsection{EVCI - Charging Infrastructure Developments}
The details of the technical advancements in the EVCI from grid to battery are found in \cite{rivera2021electric}. For fast charging, extensive cooling of the battery is mandatory, during and before charging of the battery to prevent battery aging. During this process some of the parts of the EV exceed the on-board charging power limits, these are to be branched from the charging power. Single-phase AC chargers are present at home and office parking places. The three-phase AC moderate charging stations cover up to a power of 25kW, while the fast DC charging stations are fast covering up to 400kW, and are possible to project as high as 900kW.

Japanese automakers and industries promoted "Charge de Move"  (CHAdeMO) which began with a maximum power of 50kW which is enough for 24kWh batteries, but not suitable for batteries of capacity 40kWh or more. So the specifications were updated up to 100kWh in 2017, and in 2018, further updated to 400kWh in CHAdeMO 2.0. Finally, in the third version, it was updated to charge a rate of 900kW to reduce the charging times and also accommodate larger and heavier vehicles. This version supports the current and voltage ratings upto 600A and 1.5kV. The most recent version in the domain of field testing is CHAdeMO 3.0.1, also known as ChaoJi-2, while the forthcoming version, known as CHAdeMO 4.0, or Ultra-ChaoJi is currently in the planning phase. With the increased connectivity, the EVSE network is integrated with the Internet of Things (IoT) of EVs and is represented as a necessary cyber-physical component of the contemporary grid.

\subsection{Latest Development in EV Industry}
The technologies in the battery manufacturing have advanced significantly since 2010. A typical EV battery's projected life is roughly 8 years or 160,000 miles, which is higher than earlier batteries' power and energy densities at lower prices. Because of this, the majority of mass-market EVs now on the market do possess a driving range greater than 320 km with 44kWh of average capacity. However, due to the fact that modern EVs do not get charged as quickly as anticipated when using standard 50kW chargers, this capacity development has created significant issues relating to battery chargers and charging rates. The improvements in battery technology and the usage of large size batteries may also confuse labeling the existing DC infrastructures. This results in the existing home EV owners being installed with the fast charging stations at an additional cost of around \$ 4000. The interests of EV architectures in the 400-900V range require modifications in the existing regulations. The utilization of the 800V voltage range is also employed to effectively double the charging power while adhering to the same cable constraints. Most of the standards are compatible with the EV technology voltage range of 600V \cite{rivera2021electric, bowermaster2017need, lee2018charging, jung2017power}. Even though electric vehicles are popular in many countries, they do have some technical barriers that limit their usage. The major issues are related to battery charging as explained previously. Even after the adoption of ultra-fast charging methods, due to limited charging networks and a significant amount of time spent by the EV as compared to the traditional fuel vehicle, there exists vehicle congestion at charging stations.

\subsection{Classification of EVCI}
There are many papers and articles available in the literature for the classification of the EVCI. In $2012$ the EVCI classification was based upon the type of the topology of power electronic used \cite{yilmaz2012review}. In \cite{lebrouhi2021key}, EVCI is classified into three categories based on the mode of energy transfer i.e., conductive charging systems, wireless charging systems, and battery swapping systems. The conductive charging is further divided into slow, semi-fast, and fast charging techniques. The wireless charging into static, dynamic, and quasi-dynamic charging techniques. And the battery swapping into top, bottom, side and rear swapping techniques depends upon the battery placement in the EV. In \cite{das2020electric}, the EVCIs are classified based upon the types of power flow i.e., ac or dc, and further, each category is divided into different levels that are sub-categorized by the fixed voltage and power levels. Samrat et al. in \cite{acharya2020cybersecurity} shows the classification of EVCI in combination of the cyber and physical layer.

\subsubsection{On-Board Charging (OBC)}
\begin{figure}
    \centering
    \includegraphics[scale=0.6]{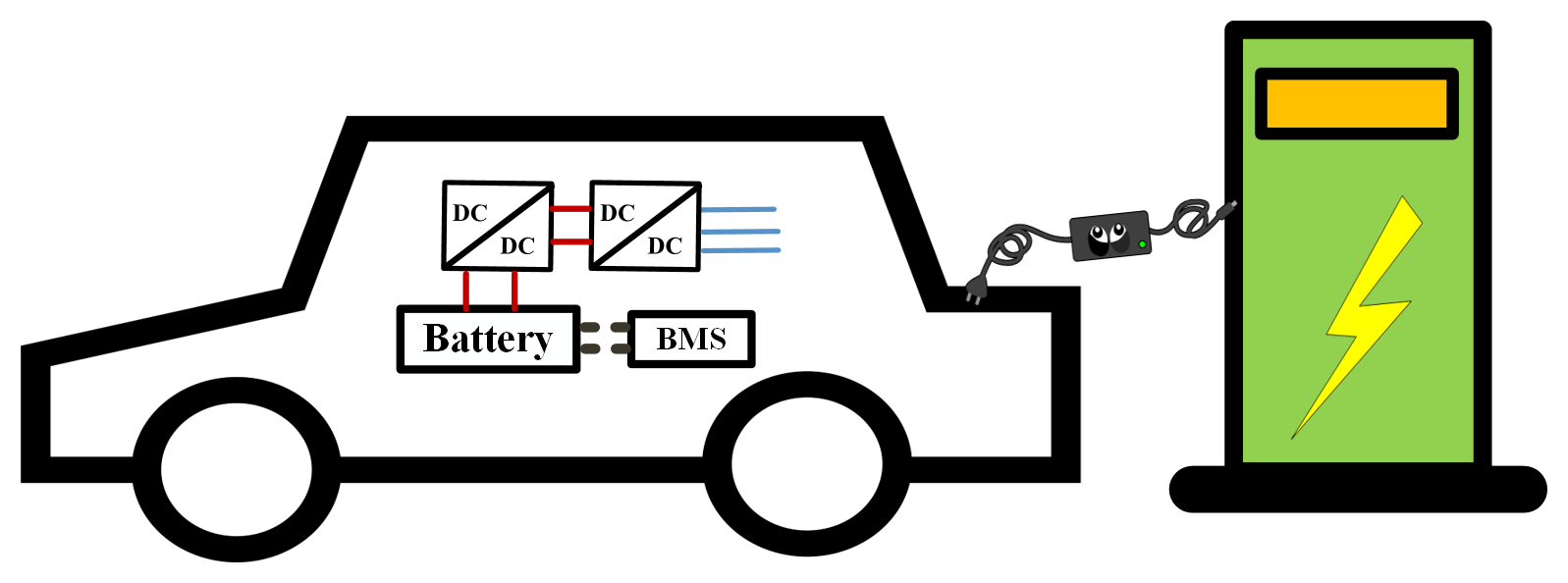}
    \caption{Schematic of On-board Charging}
    \label{On board charging}
\end{figure}

Recent study shows that home charging accounts for 72.5\% of the energy used by the Light-duty vehicles (LDV), while the fast charging option occupies the second position with $11.9$\% \cite{neaimeh2017analysing}. The schematic of the onboard charging mechanism is shown in Fig. \ref{On board charging}. Vehicle manufacturers (automobile manufacturers) have strict control on the OBC and its expansion resulting in OBCs having stringent internal communication in transferring the data than the third party off-board chargers. These are deeply concerned with safety and other monitoring programs. These chargers typically decode rather than extracting from the battery management systems BMS and might make use of other vehicle resources and offer interfaces to run third party modules like AUTOSAR, GENIVI development platform and automotive Linux. Due to accessing the high voltage and communication buses without dedicated contractors, the vehicle could stop with inconsistent behavior. So, OBCs must need to follow closely the vehicle specifications. The maximum power rating for the three-phase high-power OBC is $22$kW. The rapid surpassing of these thresholds by OBCs is compelling evidence, as it necessitates substantial advancements in conventional pathways. Additionally, because OBCs are a fixed component of the EV architecture, which is itself very package-sensitive and crammed with components, strict size, and weight restrictions force their design to focus on optimizing the amount of power provided to the battery \cite{rivera2022charging}.

\subsubsection{Off Board DC Chargers}
\begin{figure}
    \centering
    \includegraphics[scale=0.6]{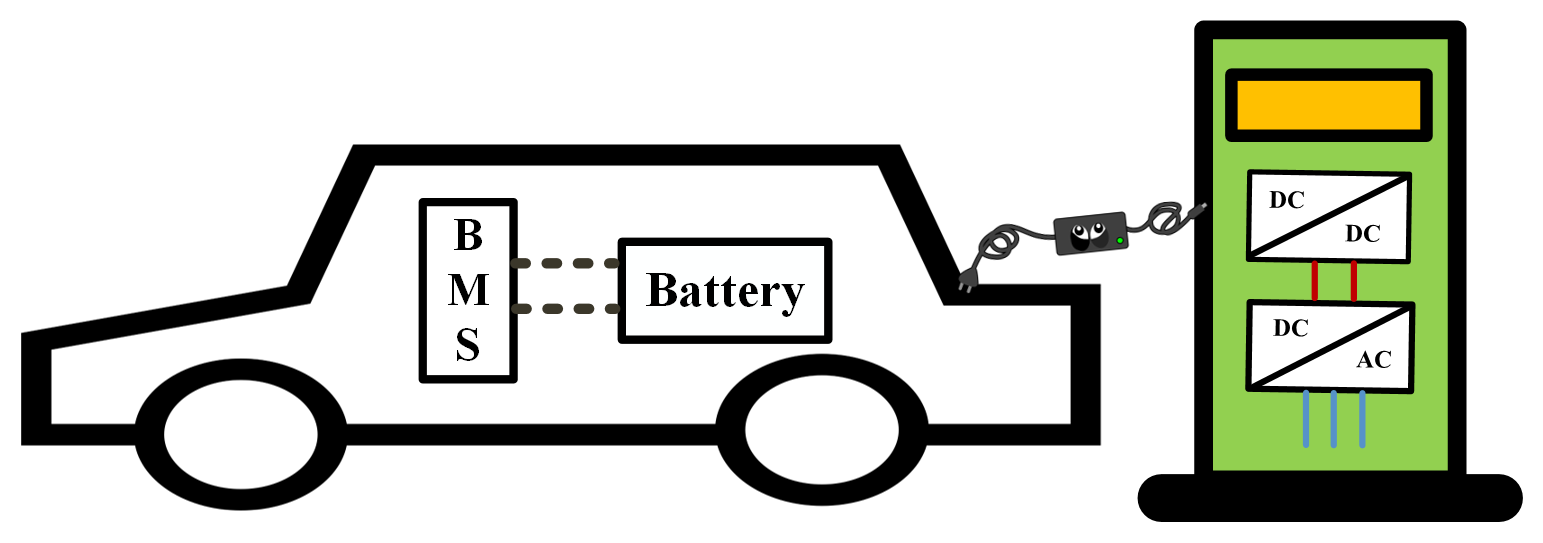}
    \caption{Schematic of Off-board Charging}
    \label{Off board charging}
\end{figure}

Fig. \ref{Off board charging} represents the schematic mechanism of the off-board charging system. These charging systems fundamentally differ from the traditional AC chargers as most of the power conversion takes place exterior of the vehicle and the battery is also exposed. This shifts the power conversion and control burden enabling the power transfer in larger amounts without the necessity of power conversion electronics on board. The current capacity of the DC charging power range is 20 to 350 kW for mid-sized vehicles. This is expected to be increased to 600 kW and is projected to develop up to 4.5MW in the future \cite{Charin}. By making the EV decide the charging power by controlling the voltage and current values, these charging stations might be made more suitable for public charging stations. EV manufacturers do not have much access to the charging stations, the EV sees the off-board chargers as a black box. So, regardless of the rating of the chargers it is required the EV must be charged at the reference values set by the BMS of the EV for safety and appropriate treatment of the battery \cite{9199855}. For any charging, it is required to power a few auxiliary loads such as the refrigerator, cooling compressor, air conditioning, and thermal management of battery, etc., For a modern vehicle, the average auxiliary load for a $12$V battery may exceed $1$kW, with several kilowatts as peaks provided the no power saving measurements are taken \cite{rivera2022charging}. 


\subsubsection{DC Chargers for Heavy and Medium Duty EVs}
Heavy-duty vehicles (HDV) and Medium-duty Vehicles (MDV), includes buses, trucks, and cargo transport vehicles, make up a category of large vehicles. This makes it possible to use larger battery capacities that may be charged at higher power levels and charge rates for keeping the charging times sensible. Many efforts have been carrying out by the standard organizations to enable the megawatt charging systems (MCSs) \cite{mcs} for conveniently enabling the drivers to charge under 30 mins \cite{blech2020project}. However, the major drawback of charging both the HDV and MDV with high charging rates and large energy capacities is that it causes a high voltage stress on the utility grids. These vehicles generally adopt for depot charging \cite{depotcharging} or opportunity charging \cite{opportunitycharging} where the former one charges the vehicle overnight with low Coulomb (C) rates and the later one uses the charging while the vehicle is moving similar to the locomotives. It distributes charging ports throughout the routes and with the use of automated charging devices (ACD) or pantographs, it quickly charges the smaller capacities with higher C rates several times throughout the day. 

Communication-wise all the above-mentioned methods adopt the different modes of communication once the EV is connected to the EVSE except ACD, as the pantograph of the system connection must ensure that the charging is allowed. Accordingly, the communication must be established before the EV and EVSE are connected so that it may include wireless mode. It is to be noted that there are possibilities for inductive charging (wireless dynamic and static charging), battery swapping system (replacing the EV batteries with the fully charged ones), and trickle charging (charging with normal current sockets of the house without any intermediate components) but it is not considered as an EVCS because it lacks key elements such as chargers and control devices \cite{costantino2023electric}.

Since the above classification deals with the positioning of power conversion systems (power electronic devices), EV charging may further be classified based upon its voltage rating and the amount of power ie., the rate at which the batteries are charged. The simplest and slowest charging method is Level-$1$ connectors, which adhere to the J$1772$ standard. In a private setting, they might use AC, which is typically $120$V. Public AC $240$V chargers that stick to the J1772 standards (includes Tesla connectors) are considered as level-$2$ plugs. The last category is the Level-$3$ that uses DC fast charging. The connectors used for charging the EVs are discussed in Section-\ref{standards and protocols}.

\subsection{Role of Communications in EVCI}
The CPPS interconnection protocol enables communication between the various integrated heterogeneous systems of CPS. The primary goal of this protocol is to provide CPPS heterogeneity at three separate levels, including functions like interoperability, policy regulation, and performance assurance. There exist three levels of interactions in the CPPS occurring at different stages \cite{3} such as:
\begin{enumerate}
    \item Generator, transformer, transmission line, dynamic load, etc., with the power system controller 
    \item Power system control and communication infrastructure.
    \item Communication infrastructure and the cyber system.
\end{enumerate}

The cyber unit in the CPPS is to perform advanced operations in the power grid like state estimation, forecasting, reactive power control, voltage control, oscillation monitoring, operations planning, stability analysis, model validation, etc., as its primary function at the healthy state of operation. 
The communication established between the cyber and physical layer are divided into two categories:
\begin{itemize}
    \item Wired Communications: Power line carrier communications (PLCC), Ethernet, CAN, fiber optic, etc.,
    \item Wireless Communications: Bluetooth, wifi, zigbee, broadcasting, local area network (LAN), web-based, smart applications, etc.,  
\end{itemize}

In an EVCI, communications are required for the information exchange between the EVs,  EV users, DSO, EMS and CSMS. Since most of the EVCI uses modern wireless communication methods but there exists some areas such as between the EVs communicating with EVCIs and the internal EV communications that follow CAN bus or PLCC protocols that are vulnerable to cyber attacks \cite{harnett2018doe}. The EVCI may internally communicate in wired or wireless mode. The wireless communication is used to performs operations such as regulating power flow from the grid and EMS by sending commands (reference and setting vlues) to the control devices of the EVCI and EMS. The EVs and the EVCI communicate with the EV users by exchanging information like pricing data, charging information, EV arriving and departure time at the charging stations, SoC, and few other values. A dark fiber base is used in \cite{nedyalkov2019attacks} for building a secure communication network for ECVI. The standards and the protocols adopted in the EVCI are discussed in Section \ref{Communication standards}.

\subsection{Battery Technologies and Role of Batteries}
The first EV was spotted driving on the road in late $1800$s, after the creation of rechargeable lead acid batteries and electric motors. But in the early 1920s, internal combustion engines dominated the EV market industry and all the EVS vanished as a result of their heavy weight (due to batteries), short trip range, prolonged charging time, and short lifespan of batteries \cite{corrigan2011batteries}. The latest trends in EV batteries and their technologies are described in \cite{IEA}. EV batteries differ significantly from batteries used in consumer electronics. They must be able to manage high power (up to $100$ kW) and large energy capacity (up to tens of kWh) while taking up minimum space, weight and cost \cite{young2012electric}. The different battery technologies used in the EV are NIMH (Nickel-metal hydrate), Li-ion (Lithium-ion), Lithium polymer, Sodium/Metal Chloride etc.,

Recent advances in the technologies of the smart grid prove that the EV batteries at the charging stations along with the energy storage system are used for supporting the grid and other ancillary services\cite{acharya2020cybersecurity} and get incentives in return for these services. The payment architecture and the methods are discussed in Section \ref{Payments}.  There are different characteristics of the battery namely capacity, energy density, charge state, specific energy, specific power, charge cycles, lifespan, internal resistance, and efficacy \cite{sanguesa2021review}. The capacity of the batteries differs from one EV to another. The integration of these batteries into EVs necessitates specific engineering design and compliance with rigorous requirements. The process involves the implementation of numerous protective mechanisms and the mitigation of various technical challenges to ensure the seamless integration of the battery within the EV \cite{zhao2022connecting}.

\begin{table*}[]
    \centering
    \caption{Classification of EV representing their battery specifications and charging duration}
    \label{evclassificaton}
    \begin{tabular}{cccccc}
    \toprule
    \textbf{Vehicle Model} & \textbf{Manufacturer} & \textbf{Type} & \textbf{Battery Capacity (kWh)} & \textbf{Range per single charge (kms)} & \textbf{Charging Time(hrs)}\\
    &&&&(approx)&(max-min)\\
    \midrule
    Leaf&Nissan\cite{Nissan}&BEV&60&243&3.5-0.5\\
    Model S&Tesla\cite{tesla}&BEV&75&650&7-0.5\\
    Model X&&BEV&100&560&5-0.5\\
    Model 3&&BEV&75&500&6-0.75\\
    i3&BMW\cite{BMW}&BEV \& PHEV&42.2&200&0.5\\
    Prius Prime& Toyota\cite{Toyota}& PHEV&8.8&40&5-2\\
    \bottomrule
    \end{tabular}
\end{table*}

\subsection{Payments-Schemes and Methods}
\label{Payments}
From 2 million in 2018 to 10 million in 2022, the global EV sales grew and hence the charging demand \cite{EVsales}. To fulfill this growing charging demand for the EVs, a competitive market situation is very certain where a number of self-interested charging stations compete with one another in this environment to increase their profits. Similarly, EVs would try to lower their charging expenses. A comprehensive pricing scheme is proposed in \cite{gupta2023comprehensive} based on the above-mentioned two conditions ie., the EVCI maximizes their profits while the EV owners strive to decrease their charging cost. The interaction between the multiple charging stations is treated as non-cooperative game theory and a Nash equilibrium (NE) pricing based on game theory is derived. The primary locations for EV charging include the workplace, the house, public spaces within neighborhood regions, and rest stops throughout long-distance traveling situations. The likelihood of charging is dependent on a number of variables, including range anxiety, practicality, the location of chargers, as well as knowledge of battery degeneration. When individuals use EVs for longer trips and have to stop frequently to recharge their vehicles, this aspect becomes much more noticeable \cite{visaria2022user}.  

There were many pricing models available \cite{pricingmodels, huang2023pricing, visaria2022user, fescioglu2023electric} and are classified as follows:
\begin{itemize}
    \item \textbf{Fixed-rate pricing model} - This pricing model is suitable for the new market player that adopts lower prices than the present competitive prices. An example of this model is shown in \cite{bhatti2018rule} where the EV charging is done at a constant rate by optimizing the energy flow between the DERs and the grid during the off-peak and the peak hours. 
    \item \textbf{Time-based pricing model} - This model is based on how long the charging period lasts. It is a structure where the EV user pays a particular amount of time for charging the vehicle per minute \cite{timeandenergypricingmodels}. Different EV models charge for different time periods. This makes it unfair to some EVs as they take longer duration to charge a similar amount of energy. The primary benefit is that the user won't have to worry about the quantity of power required to charge the EV. This method is adopted mainly in different parking lots (offices, shopping malls, movie halls etc.,) where the EV doesn't tend to leave even after the charging process is completed and thereby prevents the blocking of access to other EVs. 
    \item \textbf{Energy-based pricing model}: This model charges the customer based on the amount of energy consumed by the EV (per kWh). This method is more accurate and adopts a fair pricing scheme and also encourages the fast charging techniques. The main drawback of this model is for a typical EV driver, it is more challenging to comprehend and compute.
    \item \textbf{Hybrid pricing model}: This pricing model combines the both time based and the energy based models for ensuring the fair pricing both to the EVs and by the EVCIs. This model also prevents the access blocking of the upcoming EV by the present EVs. 
    \item \textbf{Advanced Pricing models}: These are further divided into two categories as described below:
    \begin{enumerate}
        \item \textit{Charging during off-peak/peak hours}: This price model offers the incentives and the discounts for the EVs those who charge during the off-peak hours and imposes high cost for the EVs who charge their vehicles during peak hours. Generally this model adopts dynamic pricing and the time-of-use (ToU) pricing methods \cite{9816010, cao2011optimized}. This model also helps in scheduling the charging based on prices \cite{dubey2015determining}.
        \item \textit{Pricing based on the type of EV charger}: This pricing model allocates various prices based on the different types of chargers used, the number of charging ports, and the type of charging mechanism the EV adopts. This method offers premium chargers to the users who select the fast and extremely fast charging process as compared to the other process. Besides, these models also set high priority for the EVs that are ready to pay the premium rates to avoid waiting periods. 
    \end{enumerate}
\end{itemize}

\subsection{Smart EV Switchgear}
\label{Smart EV switchgear}
Switchgear is the essential component of the level-3 charging stations (also referred to as DC fast charging stations). Conventional switchgear typically consists of electrical panels that are tasked with receiving, distributing, and protecting the power equipment's machinery and wiring. Generally, these panels are produced by different manufacturers (NEXPHASE$^{TM}$ \cite{Nexphase}, ABB \cite{ABB}, Schneider \cite{Schneider} etc.,) and their installation and wiring require engineering design. These switchgear panels exist separately for different voltage levels like high Voltage (HV), medium Voltage (MV), and low Voltage (LV) switchgear. The components of the traditional switchgear are fuses, relays, switches, isolators, circuit breakers, lightning arresters, and transformers, the panels included in the traditional switchgear are the utility meter, power disconnect panel, transformer, and current transformer cabinet. These switchgear play a crucial role in controlling the EVCI. Franklin electric fueling systems \cite{Switchgearspecifications} is one of the organizations that describes the specifications or the requirements for the smart switch gear panels of EVCI. The specifications that distinguish these smart panels from traditional panels are 
\begin{itemize}
    \item Integrated cellular modern communications.
    \item Remote EV charger power cycling.
    \item EV charger crash detection or Flammable vapor monitoring capability.
    \item Remote web interface for monitoring, control, and reporting.
    \item E-stop integration capability (seamless integration of robotic systems that strategically reach its customers).
    \item Integration pest \& rodent intrusion mitigation.
    \item Flood detection with automated shutdown.
\end{itemize}

\subsection{EVCI as CPPS}
\subsubsection{Definition of CPPS}
Cyber physical power systems (CPPS) are networked designs that use communication, control, and computer resources to interact with the environment of physical power systems. Different business sectors and vital infrastructure also rely on CPPS \cite{yohanandhan2022holistic}. The main motivation of the CPPS lies in two things namely, interoperability and integration. The word interoperability refers to a multidimensional term that incorporates multiple perspectives and methods from various groups and applications. Since it fits with the heterogeneous features of the CPS, from various angles, to prevent the ambiguities concerning CPS development the interoperability is defined in various areas like system interoperability, technical interoperability, data, process, operational, and information interoperability, etc., the extensive works regarding the interoperability is discussed in \cite{gurdur2018systematic}.

Modern generating schemes, cutting-edge control methods, improvements in data transmission through open communication networks, and the development of communication network security measures through smart systems are all contributing to the field of power system's ongoing advancement \cite{mohan2020comprehensive}. With the inventions of applications in computations and communication technologies, coupling the traditional power systems have been transformed into CPPS \cite{zhao2010cyber}. The conventional grid is modernized by the CPPS technologies and is also referred as the next-generation power system as it incorporates the above-mentioned technologies across the various power system levels such as generation, transmission, distribution, substation, and utilization. The real-time operation of the CPPS relies on the communication technologies that serve as a backbone of the CPPS.

A new area of research and development was made possible by the rise in demand for intelligent electric vehicles (EVs) and plug-in electric vehicles (PEV). Both in public and private spaces, there are now a significant number of EV charging stations. Several organizations must interact with these EV charging stations securely and effectively. By $2040$, it is anticipated that one in three vehicles will be electric worldwide \cite{garofalaki2022electric}. The integration and generalization of the service offered by the independently operating charging stations depend on numerous standards and shared infrastructures for EV charging. This integration makes the EVs easier to obtain entry in the seamless operation of various frequent charging EV stations with proper collaboration. The smart EVs are further be integrated with the smart grids which makes the system more complex and the CPS more heterogeneous \cite{bhargavi2020integration}. The standards and the protocols adopted by the EVs and the EVCI are elaborated in further sections.

\section{Standards and Protocols used in EVCI}
\label{standards and protocols}
Recent days, the researchers working in the field of CPPS have concentrating more on the stability analysis of the CPPS from control system point of view. But, it is necessary to analyse the electric power systems as an integration of physical and cyber systems. As the cyber system consists of communications, controlling and computing parts these should meet some standards and follow protocols for the seamless operation of the systems. These standards and protocols were given by some organizations namely, the Institute of Electrical and Electronic Engineers (IEEE), American National Standards Institute (ANSI), International Organisation for Standards (ISO), Society of Automotive Engineers (SAE), and International Electrotechnical Commission (IEC).

The European network of cyber security proposed several security standards for EVCIs \cite{ENCS}. Security is required for both the acquisition of EVCS and communication between the EVCS operator and the power grid operator. These standards outline message encryption for secure communication, access control, designing EVCS with future security in mind, as well as monitoring and managing system security. The US DoE, Homeland Security, and Transportation addressed the cybersecurity issues with smart EV charging \cite{harnett2018doe, lightman2020symposium}. Few security features recommended in \cite{ENCS} have been added by the US Department of Transport \cite{harnett2019government} and the US National Motor Freight Transport Association (NMFTA) \cite{nmfta, XFCcyberthreats}. One of the key documents pertaining to the cybersecurity of EVCS was published in July 2022 by the US, DoE with assistance from National Technology and Engineering Solutions of Sandia \cite{o2023sample}, to safeguard the EV charging ecosystem, by providing electricity, security, and automotive industries some recommendations based on research.

H.S.Das et al. of \cite{das2020electric} has segregated the standards related to EVCI into three categories - EV charging standards, EV grid integration (EVGI) standards and safety standards. The ISO works on standardizing the EV as a whole while the other organisations work at component level. During discharging in the charging station, the EV acts as a DER which indicates that the requirements for grid integration standards of DER also apply for the EVGI and mostly all the above mentioned organisations including National Fire Protection Association (NFPA) \cite{nfpa} and National Electric Code (NEC) defines the safety and grid integration standards.

\subsection{IEC Standards}
The IEC is a British organization that creates standards for technologies that are linked to electrical, electronic and other related fields.
\begin{enumerate}
    \item IEC61851 - It covers the standards for the operation of EV for conductive charging systems and is applicable to both off-board and on-board charging systems with supply voltages of 1000V AC and 1500V DC. The initial version of this standard is sorted into three cases depending upon the charging cable attachment and placement. Upgrading to IEC61851-21 has two versions the former explains about the electromagnetic compatibility (EMC) standards of onboard charging and the later one for the off-board charger EMC requirements for conductive charging. The standards required for fast DC charging are described in the IEC61851-23 version and for this the communication between the charging management systems (CMS) and the EV controller and the EVSE.
    \item IEC61890 - It is applicable to the wireless power transfer (WPT) based charging system and also to the WPT systems combined with onsite storage systems.
    \item IEC62196 - This standard has three sections, the first one contains general specifications for EV connectors, including plugs, socket outlets, vehicle couplers, and vehicle inlets. The second is used to standardize the types of mains in the connecting system namely, types 1,2 and 3 to the modes 1, 2 and 3. In the third section, the detailed descriptions of the specific designs for vehicle connectors and inlets tailored for DC charging in mode 4 for electric vehicles \cite{iec62196}. The types and modes are discussed elaborately in Table \ref{types of connecting system} \& Table \ref{modes of connecting system}.
\end{enumerate}

\begin{table}[]
    \centering
    \caption{Various types and modes of connecting EVs to charging stations.}
    \label{types of connecting system}
    \begin{tabular}{p{2cm}p{5.3cm}}
    \toprule
    \textbf{Type} & \textbf{Connector description}\\
    \midrule
    Type 1 & AC connector with 3 and 7.4 kW, 16A, and supports only single-phase power supply.\\
    Type 2 & AC connector with 3 and 43 kW, 16A for single-phase and 63A for three-phase power supply. \\
    Type 3 & 62.5 kW and can reach 125 A, yet the revised CHAdeMO 2.0 specification allows for up to 400 kW.\\
    \bottomrule
     \end{tabular}
    \end{table}

 \begin{table}[]
    \centering
    \caption{Various types and modes of connecting EVs to charging stations.}
    \label{modes of connecting system}
    \begin{tabular}{p{2cm}p{5.3cm}}
    \toprule
    \textbf{Mode} & \textbf{Socket description}\\
    \midrule
    Mode 1 & Standard socket outlet for home charging.\\
    Mode 2 & Standard plug outlet for home installations with a unique cable that integrates a power control and protection system.\\
    Mode 3 & Dedicated charging system (EV charger) with a dedicated circuit that incorporates control and protective features. cable with a pilot wire included.\\
    Mode 4 & Dedicated DC EVSE, for fast charging equipment.\\
    \bottomrule
\end{tabular}
\end{table}

\subsection{SAE Standards}
SAE International is a global professional organization and standards development body for the engineering industry, covering a wide range of topics related to the design, testing, manufacturing, and performance of vehicles and their components. These standards are intended to establish consistent engineering practices and specifications to ensure safety, reliability, and performance in various applications.
\begin{enumerate}
    \item SAEJ2293 - It signifies the charging needs of both on board and off board charging systems, dividing into two sections one for the power requirements along with identifying EVSE location as optional and the other for the communication and network design \cite{j2293-1, j2293-2}.
    \item SAEJ1772 - It explains the ratings of the equipment like circuit breaker along with the charging voltage and current ratings \cite{das2020electric}. It also signifies the levels of voltage and current ratings for the different charging levels. Recent versions of this standard recommends about the standard for the charge connectors for conductive charging \cite{j1772}.
    \item SAEJ1773 - This standard outlines the minimal interface compatibility specifications for an EV manually connected inductive charging system of level $1$ and $2$ chargers. The software messaging requirements are present in Appendix. A and the recent versions provides the recommended software interfacing message requirements in Appendix. B of the document \cite{J1773}. 
    \item SAEJ2931 - The requirements for digital communication between EVs, EVSEs, utilities, energy service interfaces, advanced metering infrastructure (AMI), and home area networks (HAN) are established by this standard. The establishment of an EV communication network in a smart grid context is also specified. Table \ref{table2931} shows the different versions of these standards.
    \item SAEJ2836 \& SAEJ2847 - These two standards combined with SAEJ1772 enumerate the requirements for communication between EV and EVCI. The former specifies the communication requirements and the latter one for the use case and for providing the testing infrastructure.
    \item SAEJ2954 - It is the world's first WPT specification for EVs. The initial version supports up to level 2 charging but the latest version is upto level 3 (1.1 kW). This standard also incorporates autonomous charging, smooth EV parking, and payment establishment.
\end{enumerate}

\subsection{IEEE Standards}
IEEE is a global organization that develops standards for a wide range of industries, with a primary focus being electric and electronic engineering. It covers diverse areas including telecommunications, information technology, power and energy, biomedical, etc., The standards of this organization play a vital role in ensuring reliability, interoperability, and safety in various technologies and industries.
\begin{itemize}
    \item IEEE 1547 - It recommends the practices for integrating the distributed energy resources with the grid with a collective capacity of 10MVA. It covers the requirements for testing, operating, maintenance, and other safety considerations. Including these functions, the major focus of this standard describes Islanding operations \cite{basso2004ieee}. It also explains the requirements for the grid modernization \cite{basso2015ieee}.
    \item IEEE 2030 - This standard has a lot of versions, to encourage effective and quick charging between electric vehicles and DC quick chargers, the year 2015 version standard specifies specifications for the designs of electric vehicles and DC quick chargers. Additionally, it details how electric vehicles and rapid charges work together \cite{cheng2014smart}. The latest version  released in the year 2022, and the scope of this version lies in the design interface of EV that utilize the battery EV as power storage devices and DC bi-directional chargers \cite{9541322}. 
    \item IEEE P1547 - This specifies the standards for different aspects of grid integration of DERs. This standard has lot of versions released in different years \cite{9737086, 7003964,  6687189, 9721221}.
\end{itemize}

\begin{table*}[]
    \centering
    \caption{Versions in SAEJ2931 \cite{J2931}}
    \label{table2931}
    \begin{tabular}{cccl}
    \toprule
    \textbf{Version} & \textbf{Reference} & \textbf{Recent version year} & \textbf{Titles}\\
    \midrule
    J2931/6\_202208 & \cite{202208} & 2022 & Signalling communication for wireless charged EV\\
    J2931/7\_201802 & \cite{201802} & 2018 &Security for plug-in EV communications\\
    J2931/1\_202309 & \cite{202305} & 2023 & Digital communications for EV\\
    J2931/4\_202305 & \cite{202305} & 2023 & Broadband communication PLC for plug-in EV\\
    \bottomrule
    \end{tabular}
\end{table*}

\subsection{UL Standards}
UL is a recognized standards maker in the US and Canada with more than a century of expertise creating more than 1,500 Standards. UL Standards collaborates with national standards organizations in various nations to create a safer, more sustainable society as part of its expanded worldwide public safety mission \cite{Ul}.

\begin{itemize}
    \item UL2231 - This standard specifies the personnel protection for EV supply circuits and has two sections, UL$2331-1$ revised in 2021, it covers standards for grounding of different equipment and also applicable to systems where any accessible area of the charging system offers continuous current less than $70$ mA RMS \cite{2231-1}. The second section UL$2231-2$ version covers about the photovoltaic plates and inverters by providing the AC power and also the grid integration. These specifications also apply to equipment and systems for quick shutdown \cite{2231-2}.
    \item UL2594 - This standard is applicable to conductive electric vehicle (EV) supply equipment that is designed to supply AC power to an electric vehicle with an on-board charging unit with a primary source voltage of 1000V AC or less, and a frequency of 50 or 60 Hz. This standard applies to electrical vehicle supply equipment designed for non-ventilation environments \cite{2594}.
    \item UL1741 - These specifications apply to all inverters, converters, charge controllers, and interconnection system equipment (ISE) intended for use in grid-connected or standalone power systems. To supply electricity to shared loads, interactive inverters, converters, and ISEs are designed to be run in parallel with the grid. These requirements apply to both power systems that combine other alternative energy sources with inverters, converters, charge controllers, and  ISE, in system-specific combinations, and AC modules that combine flat-plate photovoltaic modules and inverters to provide AC output power for stand-alone use or interaction with the electric power system (EPS), commonly the electric utility grid \cite{1741}.
    \item UL62109-1 - This standard specifies the safety of power electronic converters considering the issues of electrical shocks, energy hazards, mechanical risks, risks associated with high temperatures, spread of fire from the equipment, risks associated with chemicals, risks associated with sonic pressure, risks associated with released fluids and gases, and risks associated with explosions \cite{62109-1}.
\end{itemize}

\subsection{NEC Standards}
The installation, upkeep, and operation of EV charging equipment are all covered in great detail in the NEC code \cite{NEC} for EV charging stations. It addresses issues like:

\begin{itemize}
    \item Requirements of the safety for the charging installation equipment.
    \item Grounding requirements for EV charging equipment.
    \item Specifications for protection of cables and cords.
    \item Defining the specifications for equipment protection against fault currents and over currents.
\end{itemize}

In the coming years, EV charging facilities will become increasingly important. The NEC code must be adhered to guarantee user security and the effective operation of the charging stations. The installation, usage, and maintenance of an EVCI compiled with NEC requirements guarantees risk-free, dependable, and offers the best user experience.\\

\subsection{Standards for Chargers}

A few examples of charging connectors are as follows:
\begin{itemize}
    \item \textbf{CHA\textit{rge de} MO\textit{ve}} (CHAdeMO): CHAdeMO is a standard for DC charging that enables the seamless communication between the EV and the charger. It promises a simple, fast, and robust charging experience to all EV users. Almost 50,000 charging points are established in 98 countries. This standard enables over 500kW aiming to 900kW (600A X 1.5kV) to 1.8MW. Table \ref{CHAdeMO} displays the improvements in the different versions of CHAdeMO charging system ratings from 2017 to 2023.
    \item \textbf{Combined Charging System} (CCS): It specifies a single connector arrangement that has room for a Type-1 or 2 connector and a two-pin DC connector that supports charging at up to 200 amps on the vehicle side. The 2011 vehicle incompatibilities, which were mostly due to a lack of hardware standards at first, are diminished over time. These inconsistencies were addressed in 2014 with the introduction of the first and only open charging system, the combined charging system (CCS), supported by the CharIn organization. The CCS recommends a general solution that addresses high power DC, single-phase and three-phase AC. Initially, in 2016 this system's objectives were defined for chargers up to 200kW with a voltage range of 200-1,000V. The recent advances in the CCS permit a maximum power of 400kW for mid-ranged EVs and cover a 1.5kV and 3kA freight EVs \cite{rivera2021electric}.
    \item \textbf{Guobiao Standards} (GB$/$t): Manufacturers of EVs in China and India use the DC fast chargers described in the Guobiao (GB/T) 20234.3-2015 standard, which was revised in 2015 and first issued in 2011 by the China electricity council. This standard provides the controller area network (CAN) bus communication interface, which is identical to CHAdeMO, for DC charging of batteries with maximum ratings of 950V and 250A. The 2018 revision of the standard preserved this CAN communication platform and used in the ChaoJi charging standard of 2020 version \cite{rivera2021electric, boyd2019china}. 
\end{itemize}

\begin{table*}[]
    \centering
    \caption{Development in CHAdeMO technologies \cite{chademo}}
    \label{CHAdeMO}
    \begin{tabular}{p{2cm}p{7cm}p{5cm}}
    \toprule
    \textbf{Year} & \textbf{Specifications} & \textbf{Version}\\
    \midrule
    2017 & 100kW continuous power 150-200kW peak power & Medium charging\\
    2018 & (400A X 500V) & Project ChaoJi started\\
    \\
    2020 & 350-400kW (600A and 1.5kV) & ChaoJi/CHAdeMO 3.0 started\\
    \\
    2022 & IEC (62196-3, 3-1 and 68151-23) coupled & CHAdeMO 3.0.1/ChaoJi-2 is planned to release in April\\
    \\
    2023 & To be announced & CHAdeMO 4.0/Ultra-ChaoJi is planned\\
    \bottomrule
    \end{tabular}
\end{table*}

\subsection{Communication Standards}
\label{Communication standards}
To make it easier for different entities to communicate and exchange data, communication protocols offer a set of rules and principles. For the purpose of assurance, and compatibility across the various systems, a protocol would describe the interface between two or more interacting entities \cite{ferwerda2018advancing}. A variety of protocols and standards are used to control network communication in an EV charging system, also known as a Plug-in electric vehicle (PEV) network, because several entities must connect in an efficient and secure way. Various communication protocols that are associated with different EV eco-systems categorized in \cite{metere2021securing} are as follows;
\begin{itemize}
    \item \textit{Front End Protocols} - These protocols establish the connection between the EV and the EVSE along with the requirements such as plugs, charging topologies, safety, and bi-directional power flow etc., Few examples that define these standards are CHAdeMO, CCS, ISO-15118-20.
    \item \textit{Back End Protocols} - It emphasizes the communication and cybersecurity requirements that define the charging point operator and a third party. Typical examples consist of open charge point protocol (OCPP), IEC631000, Open automated demand response (OpenADR), EEBUS, and IEEE2030.5.
\end{itemize}

All the charging ports communicate with the EV by adopting the CAN (Controller Area Network)bus protocols. It allows internal communication between the different parts of an EV without a central processor. It uses electronic control units (ECU) for the message transfer in an unencrypted form. modern EVs opt for the telematic control units (TCU) for the security guarantee by internet connection. Another front-end system protocol is used in ISO 15118, it differs from the CAN bus in the way that it establishes the requirements for physical and data link layers i.e., it specifies the requirements for the wireless data transfer in charging applications.

OCPP is the most accepted standard which is in operation and installed in more than 65,000 companies and is supported by open charge alliance (OCA). The first foundation for OCPP was liad by the Dutch foundation ElaadNL for supporting the communication between the backend systems and the charging points \cite{dutchelnaad}. It defines the communication protocols between the chargers and charging station management systems. Table \ref{ocpp versions} shows the hierarchical versions of OCPP released. The most recent version-2.0 provides encrypted upgrades to the firmware, security logging and event reporting, authenticating, and secure communication safety profiles.

\begin{table}[]
    \centering
    \caption{Different OCPP versions}
    \label{ocpp versions}
    \begin{tabular}{cc}
     \toprule
     Version&Year\\
     \midrule
     OCPP version 1.2 & 2011 \cite{30}\\
     OCPP version 1.5 & 2012 \cite{30}\\
     OCPP version 1.6 & 2015 \cite{14}\\
     OCPP version 2.0.1 or 2.0 & 2018 \cite{13}\\
     \bottomrule
    \end{tabular}
\end{table}

The communication between most of the EVCI entities such as aggregator, chargers, EVSE etc., were defined by the IEEE$2030.5$ protocol \cite{8608044}. Security is its top priority. It provides certifications that are unable to be changed or revoked and must be kept private. Hence these standards are opted for a smaller area where scaling is not required. OpenADR (Open Automatic Demand Response) is the most applicable protocol in the context of data and signals exchange \cite{augello2022coexistence}. The EEBus public specification is free, but the underlying implementation (i.e., the code) is not \cite{metere2021securing}. EEBus is a set of protocols that have been used by several corporations in order to integrate communications in the IoT devices. They are particularly concerned with the data structure and communication transmitted between entities. They define multiple protocols at various communication layers.

\section{Cyber Security}
\label{Cyber Security}
Despite the existence of numerous standards and protocols designed to enhance the security of EVCI, various potential threats and vulnerabilities continue to persist. Recent research reports state that several intentional cyber attacks are being targeted against CPPS, severely impacting the stability and performance of power system operation and control. Due to the essential nature of the CPPS's functions, they are the main targets of cyberattacks intending to sabotage them which require proper measures to be taken to prevent the system affecting from these attacks. This practice of safeguarding the system by implementing appropriate measures is known as cyber security. Due to the growth in cyberattack events, cyber security is also evolving rapidly. The main concepts of cyber security include objectives, impacts, requirements, and challenges.

\subsection{Objectives}
\label{objectives}
The fundamental objectives for analyzing the cyber-security model of the CPPS are confidentiality, integrity, and availability (CIA). The use of security related policies, standards, and guidelines are required to verify the security of the EVCI's communication infrastructure. Since the EVCI grid integration is similar to the integration of the distributed energy systems (DER), the AAA framework, which stands for authentication, authorization, and accounting, is used to handle the CIA of the EVCI's cyber security. This section discusses the CIA model in its standard configuration \cite{vosughi2022cyber, trevizan2022cyberphysical,gunduz2020cyber,alkatheiri2021cyber}.

The protection of data from unauthorized access or disclosure is termed confidentiality. It means that the information access must be provided only by authorized people so the authorized users are unable to get the access. It is one of the most prominent issues for users and contains privacy. Mechanisms such as access control policies and encryption are common ways of enforcing confidentiality. Additionally, access control generally governs how entities are authorized to carry out operations. To enforce access restrictions according to distinct permission tiers, individuals are frequently categorized into various groups for the purpose of authorization control \cite{lai2017cyber}.
 
The tampering of data by anyone or anything must not be entertained. All types of data must be verified as being accurate and unaltered. Therefore, it is forbidden to update the data in an unauthorized or undetectable way. Integrity is the defense of data against unlawful erasure and modification. Integrity also refers to upholding and guaranteeing the veracity of the data. A safe real-time monitoring system for smart grids is made possible by integrity. The integrity of a system is ensured through a variety of procedures and methods, including the provision of mutual exclusion mechanisms, error detection, file system correction, cyclic redundancy checking (CRC), hardware RAID (redundant array of independent disks), and hash functions (employed to transform data of variable sizes into consistent).

Providing qualified persons with dependable, ongoing access to information is referred to as availability. The appropriate analysis, design, implementation, redundancy and configuration of a system or network generally improves availability. It has to do with having the ability to access resources or obtain information. It serves to safeguard the information system against malfunction. Information may be distorted, blocked, or delayed via availability assaults. Generally speaking, preventing denial of service (DoS) attacks that cause blackouts is necessary for data availability. The various types of attacks are discussed in Section \ref{Types of Possible Attacks and their Classification}.

Along with the CIA triad, non-repudiation is the one that has the necessity to accept data when the communication is permitted or legal, is another idea mentioned in the literature \cite{trevizan2022cyberphysical}. The data received by the assets should be denied at a later stage and this is termed as non-repudiation.

\subsection{Impacts - Technical and Economical}
The CPPS faces a serious security risk from cyberattacks. According to recent studies, the effects of cyberattacks on CPPs are growing. System stability, i.e., how the attackers' disruptive activity might affect system stability by causing cascade failure, and the economy, i.e., how the attackers make money, are the two main effects of cyberattacks. The impacts and consequences \cite{du2022review} of cyber attacks are as follows:
\begin{itemize}
    \item \textbf{Cascading Failures:} As a consequence of the interdependence among components within the power system, the failure of one piece of equipment has the potential to trigger subsequent failures in a cascading fashion, leading to cascading failures. The interaction between the physical and cyber layers of the CPPS has improved the functionality of the power system. However, due to the characteristics of such interconnected systems, these are susceptible to breakdowns, natural calamities, and notably cyberattacks. The criticality of the blackouts in the power system due to cascading failures is discussed in \cite{nedic2006criticality}. For example, a cyber attack on the Ukrainian power grid on December 23, 2015, caused a power outage that affected 225,000 people. \cite{whitehead2017ukraine}.  
    \item \textbf{Impact on Stable Operation:} The reliable operation of electricity systems may be impacted by the cascading failures brought on by cyberattacks. When cyberattacks related to false data and information are successfully performed on CPPS, it affects the steady functioning of the system. These fake measurements transmitted due to cyberattacks will lead to the control parameters taking wrong setting values that may compromise the overall stability.
    \item \textbf{Economic Impacts:} The quality of power supply depends upon its continuity, and reliability etc., Since the cyberattacks may lead to the cascading failures of the power system, the economy could be negatively impacted by the cascade failures brought on by cyberattacks, which is another major concern. The majority of attackers have economic objectives, such as pursuing their own interests or working for adversarial nations to stifle the economic growth of other nations. The economic impacts of the cyber security are as follows:
    \begin{itemize}
        \item Decrease in the economy of the power sector units and the generation companies (including DERs) due to energy theft. This was achieved by the attackers by directly modifying the data in their own AMI (adopted by smart grids) for economic benefits and also by changing the pricing scheme of the system. Both circumstances will result in illegal profits for the attackers. For these types of attacks, the attacker doesn't need to have the domain knowledge of the power system. 
        \item Grid topology changing attacks and load disruption attacks by creating unnecessary load shedding, cascading blackouts, and few power disruptions. These attacks may also lead to a decrease in profits because of the interruptions in the power supply.
    \end{itemize}
    Furthermore, the detailed investigation of the economic impacts of the cyber attacks is done in \cite{avraam2023operational}.
\end{itemize}

Although data market strategies are relatively novel concepts, data-driven applications have already gained a significant foothold in the energy sector.  Demand response, short-term and long-term load forecasting, electricity price forecasting, RES generation forecasts, and study of power system faults and failures are some of the factors impacted by cyberattacks on the energy markets.

\subsection{Requirement of Cyber security}
Authorization, authenticity, and accountability (AAA) are the basic requirements of cyber security as mentioned in the above Section \ref{objectives}. Apart from these the additional requirements are privacy, dependability, survivability, and safety criticality. This section explains the various requirements of cybersecurity along with their inter-dependability.

Dependability refers to the system's ability for accurate and on-time delivery services time while avoiding serious internal flaws. Dependability guarantees that services are provided even when there are internal problems. The fundamental characteristics of dependability include availability, reliability, maintainability, safety, and security. Important steps to maintaining dependability include fault tolerance, fault forecasting, fault avoidance, fault detection, problem removal, maintainability, and safety engineering \cite{alcaraz2012analysis}.

The ability of a system to perform its tasks, preventing malicious, intentional or unintentional faults on time is called survivability. The goal of survivability is to offer services both in the presence of harmful purposeful behavior and outside flaws. The attributes of survivability are dependability, availability, fault tolerance, safety, and security. The important measures taken for achieving survivability are as follows: isolation of affected areas, use of heterogeneous security technologies in network design, maintainability, redundancy, security policies, accountability, authentication, authorization, non-repudiation, and cryptographic services.

The provisions for accountability and audit are traceable and recordable. There are typical and commercial consequences for the violation of accountability. Accountability operations determine who is responsible during a security problem.

Safety criticality is a significant variation of safety, which is a crucial security requirement in CPSs. Systems that could potentially result in severe results because of the existence of some unforeseen circumstances, such as earthquake, flood, or tsunami, may cause significant physical damage, human injuries, or even fatalities.

\subsection{Challenges in Cyber Security}
Security risks are often calculated as the product of the impact of a particular scenario and its likelihood of occuring. This description works well for random sources of system failure, but there is debate about whether it accurately describes situations when risks are fueled by adversary behavior, like cybersecurity. Therefore risk assessment should focus more on the consequence of cyberattack not on the probability of the occurrence \cite{trevizan2022cyberphysical}. Due to the rapid growth in the EV and the charge points, they collectively represent a huge increased demand for electricity. This overloading might be avoided by opting for smart charging. The stability and security of the wider power system could be at risk if charge point cyber security is not properly controlled. Thus, having strong cyber security is essential. A significant number of charge points could be hacked to cause harm at different levels. local overloading of the distribution grid and a disruption that causes system balance \cite{van2022cyber}. In this context, the difficulties faced in the implementation of cyber security in the EVCI are as follows:

\begin{itemize}
    \item Implementation of a secured communication infrastructure at each and very charging station including home charging stations that charge at very low power levels.
    \item Securing the public charging stations as they are accessible by everyone.
    \item Proper load scheduling to prevent overloading and any other grid disturbances.
    \item Since the EV charging stations didn't face any major cyberattacks, many charging infrastructures are still adopting to the older versions of protocols and standards which are vulnerable to many points.
    \item Smart grid challenges that include grid integration and other automaton systems.
    \item Proper examination of the reliability of internal EV components \cite{euceda2023cybersecurity}.
\end{itemize}

\section{Vulnerabilities and Attacks Classification}
\label{Vulnerability discussion}

\subsection{Vulnerability}
The phenomenon of the weakness in a CPS that is capable of being exploited for cyber attacks is termed vulnerability. The types of equipment that are being exploited by the intruders are called vulnerable nodes. A vulnerability assessment method based on two-stage optimization is presented in \cite{akbarian2023vulnerability}. The first stage is formulated from the attacker's perspective and the second stage from the EV owner's. In the first stage, the attacker tries to maximize the load shedding, and the EV user response is revealed in the second stage by using the detected charging stations' behavior, a satisfactory function is introduced for showing the customer behavior to the cyber-attacks. The loss of load is the useful vulnerable index caused by cyber attacks. To evaluate the impact of cyber vulnerability, the modeling of the interaction between the cyber system and the physical grid is necessary. Anomaly in charging capacity may cause abrupt changes in the state of charge (SoC), which could lead to serious safety concerns. Statistics show that there were 124 documented EV fire accidents in China in 2020, with $23$\% of the fires having a charging process origin \cite{wang2021data}. To analyze the vulnerability of EVCSs to cyber assaults on the communication between EVCSs and electric utilities, a risk assessment methodology for large-scale EVCSs was created \cite{wang2019electrical}.

\subsection{Types of Possible Attacks and their Classification}
\label{Types of Possible Attacks and their Classification}
The attacker's motivation presented in \cite{akbarian2023vulnerability} is to change the charging prices by introducing FDIA, as there will be high prices in the off-peak periods and subsidiary prices during peak times. Intruders are able to modify the data of the system so as to modify the observability and controlability of the system. 

\begin{figure}
    \centering
    \includegraphics[scale=0.45]{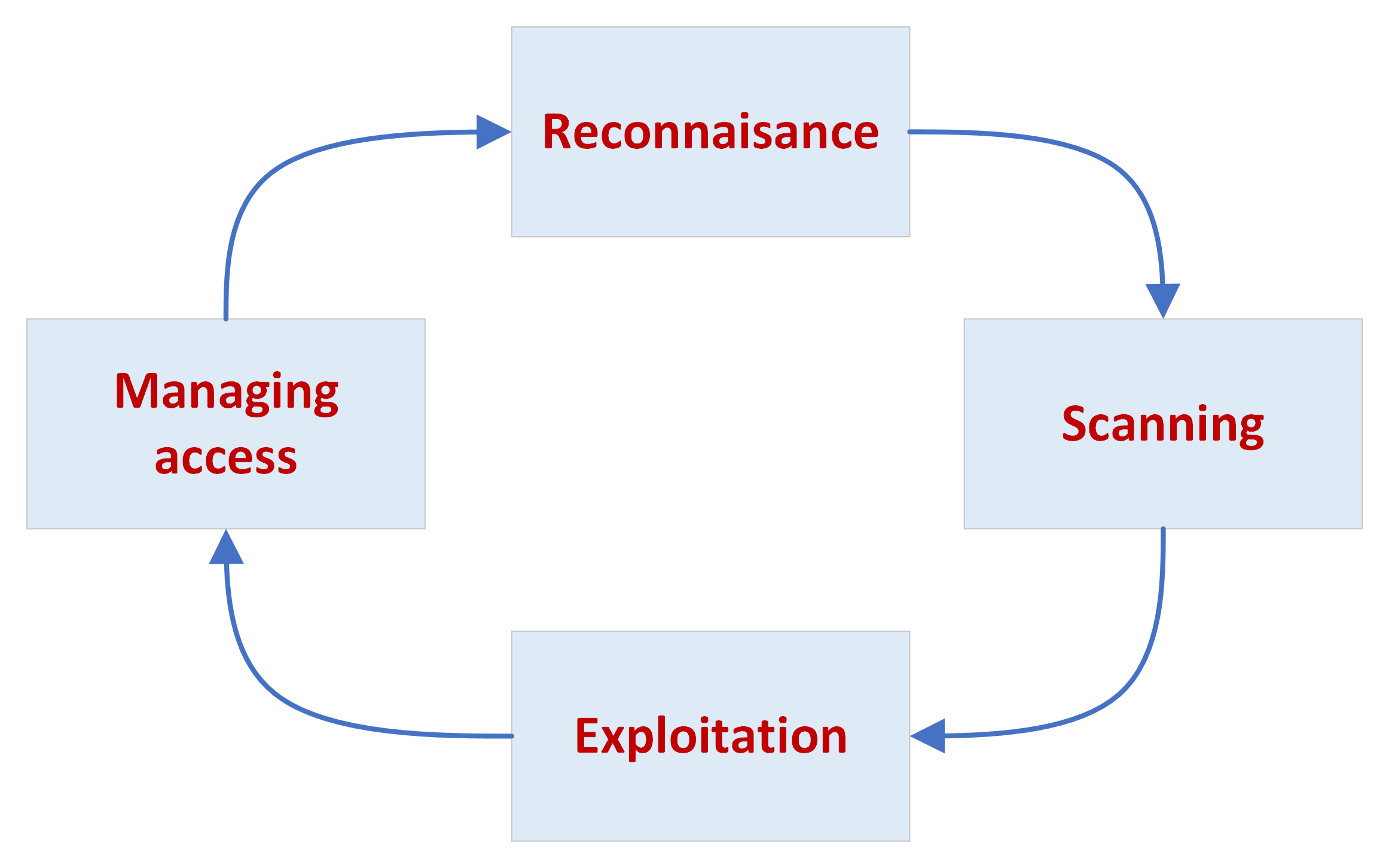}
    \caption{Cyber intrusion spiral}
    \label{attacking cycle}
\end{figure}

All intruders employ consistent execution methods, such as reconnaissance, scanning, exploitation and maintaining access as shown in Fig. \ref{attacking cycle}, regardless of the targeted system, whether an EVCS or IT \cite{el2018cyber, knapp2013applied}. Hackers gather data about the target during the reconnaissance phase using traffic analysis tools or social engineering (SE). By inspecting the IP address, open ports, and services using each open port, the scanning phase locates system vulnerabilities. In the exploitation stage, hackers try to take control of the system by taking advantage of EVCS component weaknesses. In this stage, a variety of attacks may be launched, including jammer attacks, DoS attacks, replay attacks, MITM attacks, malware attacks, and ransomware attacks. The final phase involves hackers launching different attacks—primarily viruses, Trojan horses, and backdoors to get unrestricted access to the EVCS system \cite{basnet2020deep}.

The authors of \cite{nedyalkov2019attacks, fraiji2018cyber} classifies the type of possible attacks in the EV charging stations and electromobiles based on the information exchange and also the attacks that occur the vehicle sensor network \cite{rouf2010security, sun2017attacks, petit2015remote} (like jamming attack, FDIA, DoS) as follows:

\begin{itemize}
    \item \textbf{Eavesdropping}: It is a kin to listening the communication between the control center and the charging station. It is accomplished by reading the content of the packets that have been captured from the information flow \cite{zhong2022eavesdropping}. Unauthorized interception of private information related to the vehicle, including its position and payment details, through eavesdropping, can lead to privacy breaches and potential privacy attacks.
    \item \textbf{False Data Injection Attack (FDIA)}: This attack is perhaps considered as more dangerous than any other attacks because of its stealthier nature and it enables the attacker to disrupt the normal operation\cite{liu2022false}. The infected or hacked charging station may communicate with the EV or control center with false information without the control center operator's knowledge. The effects of the FDIA on the EV markets such as biased demand schedules, incorrect forecasts of demand in EVCS, and manipulating the payments for the EV Soc data owner are discussed in \cite{acharya2022false}. 
    \item \textbf{Modification Attack}: In this type of attack, an attacker is capable of able to modify or alter the information \cite{konstantinou2015impact} received from or sent to the EVCS, EV, cloud server. The crucial information that might be modified in the EVCI systems are battery temperature, State of Charge (SoC), charging current, departure time of EV etc.,
    \item \textbf{Denial of Service (DoS)}: An attacker floods the control center with lot of messages to prevent it from gathering data from the sensors. The main intention of this attack is to prevent the user from accessing a resource. For instance, a sensor or user command is unable to enter the control room, when the number of sources launches an attack \cite{hussain2003framework}.
    \item \textbf{Replay Attack (RA)}: These are the type of data integrity attacks that are implemented by performing a prior disclosure attack for collecting sequences of data from the compromising sources and thus it replays the collected data till the end of the attack\cite{ghiasi2023comprehensive}.
    \item \textbf{Spoofing Attack}: It is the impersonation of someone or something in order to access private data including account details, passwords, usernames, and more. Sometimes the spoofing attack may consider accessing the data related to a power supply \cite{soykan2021disrupting, kosmanos2020novel} like the voltage, current, power, charge status, and SoC, etc., 
    \item \textbf{Man-in-the-middle Attack (MITM)}: This type of attack is deployed in the device that secretly modifies or alters the information communicated between the two devices while they are connected. These attacks specifically target the data transmission occurring between two endpoints in a single network or between different networks \cite{conti2016survey}.
    \item \textbf{Jamming Attack}: The jamming attack may be conducted in order to prevent authorized sensors from connecting with the control center, as well as to stop the EV car from sending these data to the server (speed, temperature, gear location data, cruise control setting, and battery status). Additionally, the driver won't be able to assess the state of the EV \cite{su2012secure}.
    \item \textbf{Blackhole Attack}: To intercept the transmitted packets, the attacker endeavors to persuade the sender (i.e., the vehicle sender) that it lies along the most expedient route. As a result, the attacker decides not to broadcast the message and dump it. Even with appropriate buffer storage, generally, the blackhole attackers discard all communications received \cite{fraiji2018cyber}. 
    \item \textbf{Grayhole Attack}: It is also known as selective forwarding attacks \cite{fraiji2018cyber}, it is similar to blackhole attacks but more severe than it is, in the view of selective dropping of messages. The behavior of the attacker is the same in both attacks. For instance, an affected node might opt to forward all messages except for those requesting energy consumption predictions or alerts about traffic safety. The attacker tends to revoke the legitimate link between the vehicles. 
    \item \textbf{Wormhole Attack} : The intention of the attacker is to nullify the legitimate communication links. The affected entities at different geographical locations may collude in order to transfer the unauthorized data. Accordingly, any exchange of data between two legitimate sources will be passed through the attacker. The different attackers in the network are coordinated, thus exchanging information among them in order to reintroduce these messages in the network (for creating network congestion).
    \item \textbf{Time Delay Switching Attacks} (TDSA): Adversaries prompt the TDSA attacks by introducing the latency in the sensors and control loops. In the EVCI the delay in sensors may result in the loss of dynamic stability \cite{ghiasi2023comprehensive}. A grid-connected EVCI network as a part of smart grids with TDSA is typically depicted as a combined system by the action of a switch that includes the phrase ``Off/Delay-by-$\tau$", where $\tau$ indicates the random delay period that is indicated by the control signals or measuring condition. Consequently, introducing temporal delays into different dynamic system modes may stimulate instability in the electrical grid \cite{huang2022cyberattack}.
    \item \textbf{Password Handling}: Password handling attacks may occur when the exchange of passwords took place in an unencrypted format. By using the access that the intruders gained, they are able to log-in without authorization and change configuration of DERS operation \cite{carter2017cyber}.
\end{itemize}

Delay-tolerant networks are generally vulnerable to attacks like blackhole and greyhole \cite{pham2015detecting}. Even they have ample buffer storage, blackhole attackers discard all communications they receive. To avoid rising suspicion and being discovered by other nodes, greyhole attackers discard a portion of the communications they receive. As a result of the dropping bad behavior, fewer messages will be delivered overall, and the intermediate nodes that have transported and forwarded the dropped messages would have wasted their resources.

Since the EVCI has the ability to be modeled as a CPS, different cyber attacks may occur at different layers. Depending upon the layers of the network the attacks are segregated as follows:
\begin{itemize}
    \item \textit{Attacks on Cyber Layer} - These consists of attacks that occur in different areas of the cyber layer such as on cloud server, privacy attacks, communication lines, identity theft attacks, spoofing, etc.,  Although these attacks target the physical components, the impact of these attacks would be severe and affect the objectives of the cyber security (CIA).
    \item \textit{Attacks on Physical Layer} - Physical layer attacks are the attacks that affect the physical equipment of EVs, EV charging stations, storage system, and the grid. FDIAs are generally caused by changing the reference values at the various power conversion stages and the power control loops are the most common type of attacks that occur at this layer.
\end{itemize}

\section{Detection \& Defense Methods}
\label{Detection and defense methods}
\begin{figure*}
    \centering
    \includegraphics[scale=0.2]{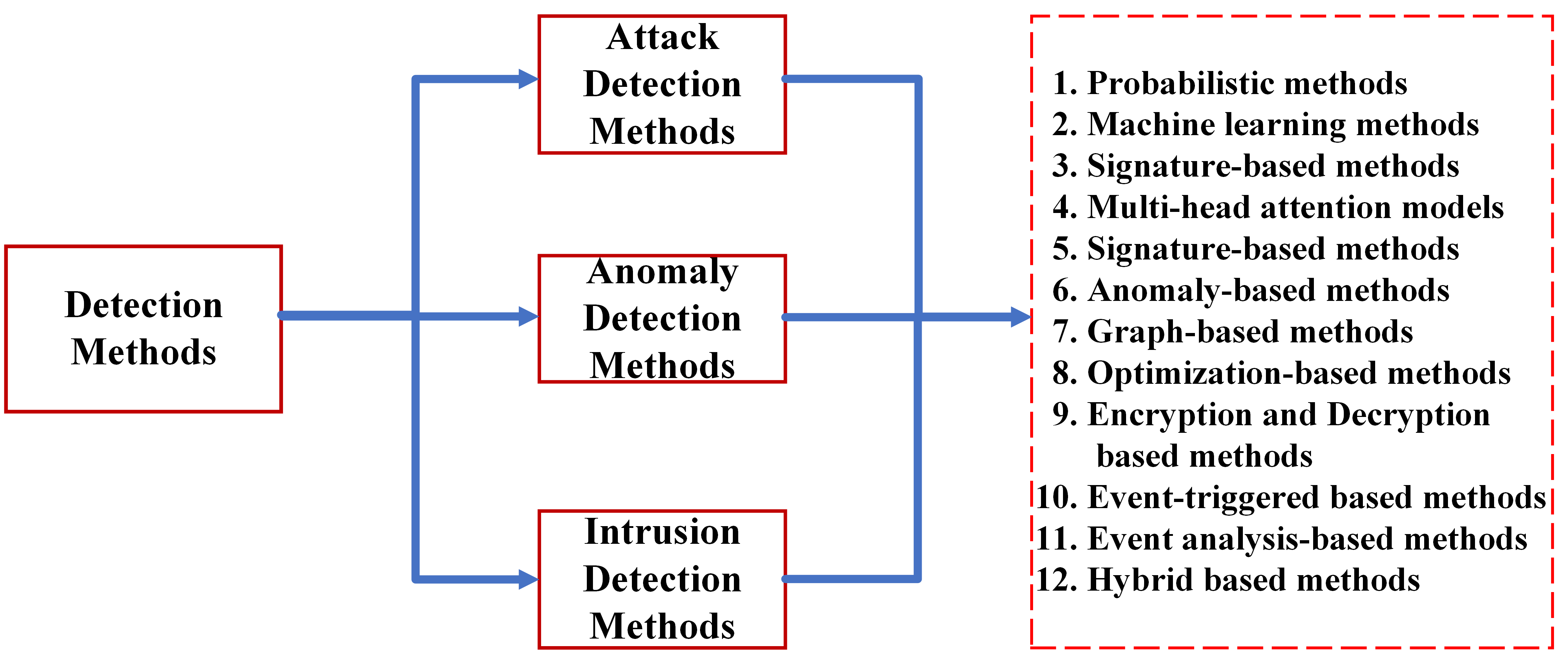}
    \caption{Classification of detection methods}
    \label{detection classification}
\end{figure*}

Cyberattacks are defined as an action or series of actions that result in a security violation. These infractions may be purposeful or accidental, and they may be committed by individuals inside the company (i.e., an insider threat) or outside it. The exploitation of vulnerabilities by malicious actors leads to the cyberattacks. The Fig. \ref{detection classification} argues about the classification of different detection methods. An intrusion detection system (IDS) is a piece of hardware or software that watches network traffic in order to separate attack data from regular data. This IDS category includes hybrid, anomaly-based, graph-based, signature-based, and machine learning IDS. 

Islam et al. of \cite{islam2022intelligent} proposed an adaptive differential privacy-based federated learning framework for building a collaborative network intrusion detection scheme model for EVCS. Within this framework, a privacy allocation mechanism, driven by reinforcement learning, is implemented at the EVCS level.

Basnet and Ali use deep neural network (DNN) and long short-term memory (LSTM)  in \cite{basnet2020deep} for the intrusion detection, where a three-layer DNN uses two hidden layers with 64 hidden units for binary classification and 128 units for multi-class classification. The simulations are performed in Python showing the results that the LSTM is superior to the DNN, with the performance metrics, used are accuracy, precision, and true positive measure. 

An event-based framework is proposed in \cite{girdhar2022cyber} for the cyber attacks on EV charging stations consisting of several processes like log acquisition, preprocessing, correlation, sequencing, analysis, and reporting that have been carried out by the investigation team. The investigation team gathers evidence and analyzes to identify the 5Ws and 1H (Who, What, When, Where, Why and How). Every time a cyberattack takes place, the attacker takes advantage of a platform by exploiting a weakness in a target component, causing an unintended event in order to accomplish their objectives.  

For anomaly detection \cite{dwivedi2023dynamopmu}, data logging must be thoroughly analyzed and suspected anomaly occurrences must be correlated. Identification of unauthorized activities and event correlation based on data and information is done by IDS where the correlation is based on spatial or temporal relationships. To enable prompt mitigation, the detection mechanism must be able to meet specific accuracy and physical constraints \cite{liu2011intruders}. The loss of load is the crucial useful vulnerability index for assessing the revenue loss and interruptions in power caused by cyber attacks.

\subsection{Consequences of Attacks}
The voltage and frequency instability caused due the cyber attacks on the EV charging stations and also the increased load demand by varying charging prices even though the distributed generation (DG) is presented in \cite{akbarian2023vulnerability}. Mohammad Ali et al. of \cite{sayed2022electric} describe the EV attack impacts on the power grid operation by simulating a power injection attack that causes frequency rise and voltage violations. To analyze the effect without an EV charging station, the same simulations were performed by replacing it with a regulated generator that shows smaller deviations as compared to the former case. Another simulated attack of the switching variety was conducted, involving the manipulation of power injection and power demand parameters, resulting in a loss of system coordination.  

\subsection{Types of Attack Detection Methods}
EVs and EV charging stations are susceptible to similar attacks that may be launched against any household load, such as altering the power demand to produce line tripping and cascading failure. Numerous articles in the literature address attacks on EVs or attacks against EV users and the power grid through the EV ecosystem. The recent comprehensive insights are provided in \cite{limbasiya2022systematic}, regarding the security and privacy-related challenges and threats confronting connected and autonomous vehicles. Analysis of several attack detection and prevention measures were also given. Attackers who take over the EV's BMS through hacked web services or malware installed on the vehicle's systems may seriously harm the vehicle itself \cite{sayed2022electric}. \cite{sagstetter2013security} deals with the attacks in EVs that involve the tampering of battery safety. The authors also examine the potential use of formal verification techniques to strengthen the security of future Ethernet/IP-based vehicle designs. The EVCI consists of numerous vulnerable components that are susceptible to different types of attacks and threats. Dajun et al. classify the cyberattacks as multi-layer, multi-type and multi-point depending upon the device or group of devices it is attacked \cite{du2022review}.

In \cite{islam2022intelligent}, dynamic optimization of the privacy budget and utility is employed to circumvent the necessity for human interventions, such as domain knowledge experts. Empirical findings validate that the privacy provisioning achieves an accuracy level of approximately 95\%.

A probabilistic bayes theorem based anomaly detection method is used in \cite{girdhar2022cyber}, but the proposed methodology is used for post attack event based analysis rather than for attack detection and the investigation reveals that due to tampering of the PCC (point of common coupling) breaker and making it trip falsely resulting in the BMS displaying the inaccurate charging and discharging schedule that are leading to anomalous status.

An intelligence driven computer network defense method was introduced where the attacker employs several attack strategies and steps to achieve their objectives. For a thorough examination of such an advanced persistent threat, these techniques are mapped with the seven steps of the cyber kill chain, which are reconnaissance, weaponization, delivery, exploitation, installation, command and control, and action on objectives \cite{hutchins2011intelligence}. 

\begin{figure}
    \centering
    \includegraphics[scale=0.2]{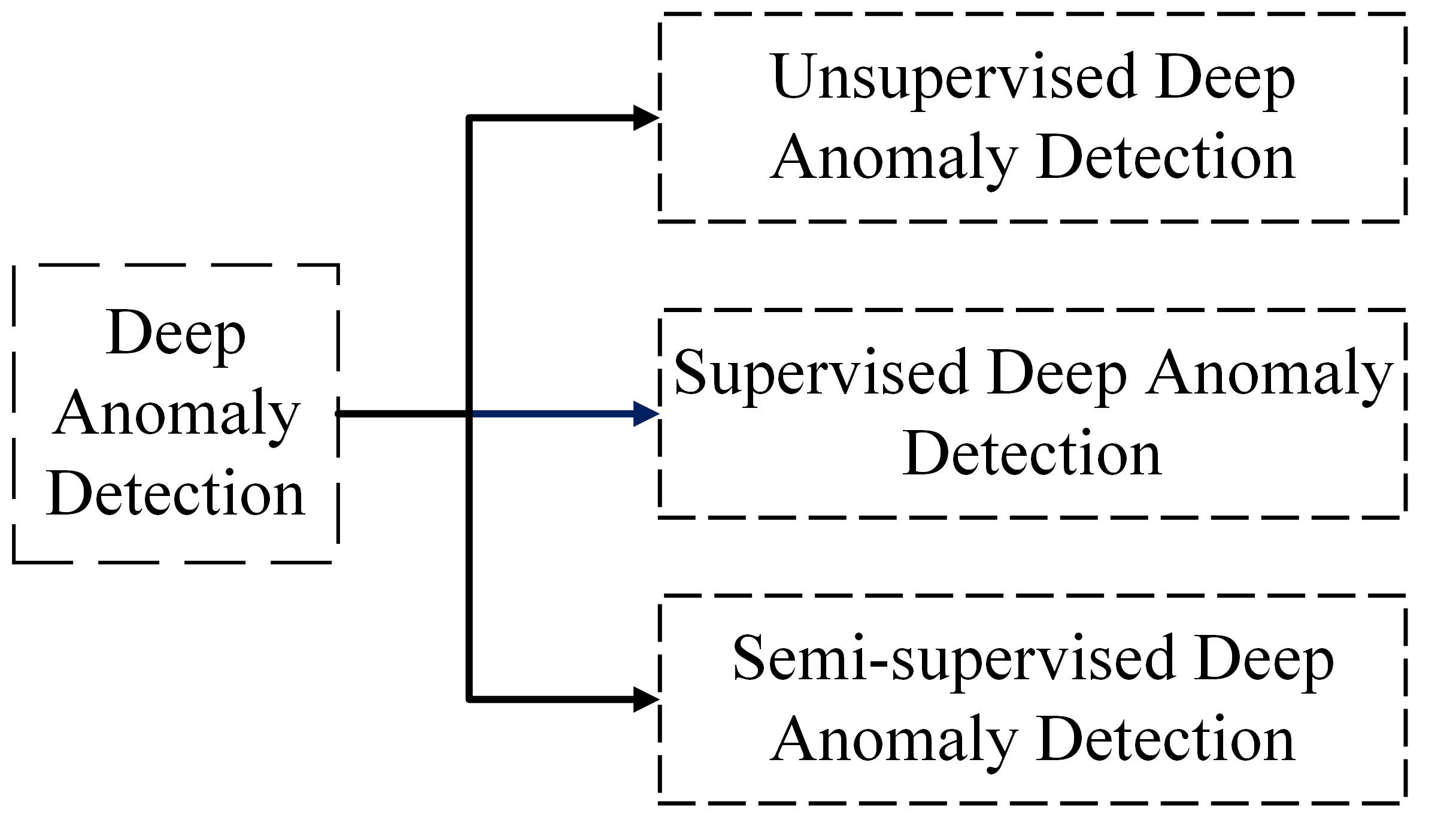}
    \caption{Deep anomaly detection framework \cite{9881073}}
    \label{anomly framework}
\end{figure}

The term anomaly refers to the outliers or the abnormal behaviors from the regular patterns that occur due to the cyber attacks and finding trends and outliers that point to unexpected or abnormal behaviors is known as anomaly detection. It is a technique for finding online invasions that necessitate data analysis and event correlation \cite{chandola2009anomaly}. A hidden Markov model (HMM) based anomaly detection method for EV charging is proposed in \cite{girdhar2021hidden} by using the STRIDE (spoofing, tampering, repudiation, information disclosure, DoS, elevation of privelage) model for designing the vulnerability assessment, threat modeling, and risk analysis. The presented model incorporates the Viterbi algorithm, in addition to HMMs, to enhance its capabilities in identifying anomalies and uncertainties resulting from attacks. This augmentation significantly improves the model's capacity for detecting potential threats. Nevertheless, the model leverages a partially observable Monte Carlo planning (POMCP) algorithm to account for the possibility that highly skilled and trained attackers may not follow the most apparent attack paths. During performance evaluation, the proposed algorithm demonstrated a false positive rate of 0.013\% and a false negative rate of 0.016\%. Furthermore, the method exhibits proficiency in detecting attacks, managing coordinated attack scenarios, making attack predictions, and executing active mitigation measures as necessary.

Jeong et al. of \cite{jeong2022electric}, were one of the first in the mathematical formulation of attack strategies aimed at disrupting EV charging systems (EVCS) by tampering with EV user data, resulting in EMS malfunctions. They devised an MITM attack, employing a bi-level optimization approach. This method involves the stealthy injection of malicious data into the system, inducing EMS malfunction and consequently elevating the overall cost incurred by the EVCS through manipulation of the EV charging schedule.
 
The integration of IoT devices within EVs, EVCS, and the utility grid heightens their susceptibility to cyberattacks. Such attacks, primarily aimed at supervisory control and data acquisition (SCADA) systems, have been identified through the utilization of machine learning classifiers, including JRipper and Adaboost algorithms \cite{hink2014machine}. However, this method exhibits certain limitations, notably a high false positive rate and computational complexities within the classifiers. Furthermore, an investigation an attack targeting the spoofing the address resolution protocol \cite{premaratne2010intrusion} targeting the SCADA systems did not yield an effective detection technique as reported by authors.

A back propagation neural network (BPNN) based detection method is proposed to identify the public EVSE incompatible demands in public EVSEs, aiming to perform switching attacks \cite{kabir2021two}. This model uses two areas with a substantial number of geographically distributed EV charging stations, both public and private. It is believed that the attacker creates an attack by directing the EVs resulting in the EVSE, to switch quickly between the charging and discharging stages in order to cause oscillations between the two areas. Such adversarial attempts are detected by using a BPNN at the CMS/CSMS of the public EVSE. The model consists of the ReLU activation function in the hidden layers to avoid the vanishing gradient and the softmax function in the output layer for generation of the binary outputs, 1 indicating the attack is caused and 0 for the no attack has occurred. Since none of the neural network models are 100\% accurate, this BPNN model mainly focuses on data of charging stations to detect the attack and the low probability of attacks may be overlooked, a specifically-tailored  H$^{\infty}$ wide area damping controller has been added to the neural network (NN) model to make sure that after a successful switching attack, inter-area oscillations are present. The design of this wide area controller is predicated on an optimization strategy, which serves as a mitigation approach to mitigate inter-area oscillations within the power grid resulting from successfully executed attacks. In this framework, the BPNN functions as a filter rather than immediately responding to the suspicious commands, introducing random delays in the execution to disrupt the attacker's coordination. Despite the BPNN's strong accuracy, it is essential to acknowledge the non-negligible probabilities of false positives and false negatives. These probabilities encompass the misclassification of a valid request as malicious (false positive) and the incorrect categorization of a malicious request as valid (false negative). The metrics employed for performance assessment include the percentage of errors, the percentage of false positives, and the percentage of false negatives.

Many researchers conclude that most of the anomaly detection systems don't fit for the anomalous data as they fit for the normal data due to over-fitting \cite{ma2021comprehensive}. To overcome the problem of insufficient anomaly data, an augmentation based anomaly detection for charging piles is presented in \cite{sun2022data}. This proposed methodology uses generative adversarial networks (GAN)- random forest (RF) where the GAN networks are used to expand the anomalous data (anomalous data enhancement network) and the RF is used for the single classification of the data (anomalous data classification network). The analysis indicates that the data with the same charging and connection time are termed normal data, whereas the same with different times are termed anomalous data. The accuracy of this method needs to be improved further. 

Kaymakci et al. of \cite{kaymakci2021energy} proposed a method for anomaly detection in energy for industrial applications. The method uses LSTM-AE (long short term memory- auto encoder) for anomaly detection. However, this article suffers from the non-holistic approach and doesn't consider the information about anomalies.

A statistical approach method was introduced in \cite{ye2002multivariate} to determine the intrusion for the multivariate time series data. This method follows the assumption that the data follows the normal distribution. $T^{2}$-test is used in this method to detect the intrusions in data. The output of $T^{2}$-test is compared with the standard F-distribution table, and the performance analysis is also compared with the $X^{2}$-test. An intelligent anomaly detection scheme is introduced in \cite{nakayama2017energy} that predicts the energy data as an indication for future failures. After observing these detections, an expert, such as an operator, provides feedback that is used to improve the outlier filtering process for further anomaly detection. Due to the adoption of efficient human feedback analysis, the performance of outlier detection is improved.

A comprehensive examination of various deep anomaly detection techniques applied to the charging data with an emphasis on deep learning is presented in \cite{9881073}. This paper doesn't consider the supervised anomaly detection methods, due to their sub-optimal performance, primarily stemming from issues related to data imbalance within the anomaly dataset. Fig. \ref{anomly framework} depicts the classification of this framework. Due to the difficulties in acquiring anomalous data, the imbalance created in the data labeled training sets for anomalous datasets is not always readily available in many applications. Because of this, machine learning typically establishes a scoring function to assess the level of anomalies in the data and ranks the data in the scoring function from the highest to the lowest anomaly scores, with high anomaly scores being indicative of anomalous data. When compared to the unsupervised deep anomaly detection model, the semi-supervised deep anomaly detection model, which also makes use of labeled data from one category, performs significantly better. The secret of semi-supervised deep anomaly detection is to maximize the usage of a large number of heterogeneous unlabeled data to enhance the training model's detection performance on computationally massive unknown datasets. Table \ref{anomly} shows the different anomaly detection methods based on deep learning.

\begin{table*}[]
    \centering
    \caption{Anomaly Detection Methods}
    \label{anomly}
    \begin{tabular}{ccl}
    \toprule
         \textbf{References} & \textbf{Detection Method} & \textbf{Summary}  \\
         \midrule
         \cite{parameswarath2023prevent}&Message Tampering & A light weight mechanism for prevention of attacks.\\
         && Computationally cheap. \\ 
         && Detects the presence of adversary who modifies the charging request messages.\\
         && Fails when the EV sends incorrect charging request\\
         \\
         \cite{mousavian2015cyber}&Worm Malware propagation & Mixed integer linear programming (MILP) is used for isolating the infected nodes. \\
         &&A response model that finds an optimum solution is determined to minimize the threat.\\
         &&This solution was not extended upto grid level.\\
         \\
         \cite{kaymakci2021energy}& LSTM & Since this proposed methodology uses a deep neural network for more accuracy\\
         && score, it is not a holistic approach.\\
         \\
         \cite{cui2019machine}&Navie Bayes Classifier&Easy to implement with low model size and works on a small amount of training\\
         &&data and converge rapidly than discriminate models.\\
         && This model fails in broken topologies and preprocessing techniques\\
         \\
         \cite{streubel2019detection}&Moving average Anomaly & Initially this model is used for the super harmonics detection.\\
         &Detection Method& It might also be used for the anomaly detection caused by the cyber attacks.\\
         \\
         \cite{li2020detecting}&Multi head attention models& High accuracy and F score, used in traffic anomaly detection at charging stations for \\
         & with CNN based network structure & industrial control systems (ICS)\\
         \\
         \cite{myers2018anomaly} & Using process mining & Detecting unusual behavior and cyberattacks by employing ICS data logs and the \\
         &&compliance checking analytical technique based on the process mining disciplines.\\
         \\
         \cite{gupta2022anamoly} & Three ensemble methods are  & Hybrid, scalable and safe. Better performance as compared with one classifier \\
         & used: random forest, decision tree & and also works better for big data.\\
         & regression and gradient boosting & It is applicable for intrusion detection systems also. \\
         & tree & Evaluated metrics indicate accuracy($99$\%), false positive rate($.15$\%) and F score ($99.03$\%)\\
         \\
         \cite{huda2018malicious} & Deep Belief Network (DBN) and & The initial model uses distinct datasets for training and testing for both the DNN and ANN.\\
         &Artificial Neuron Network (ANN) & In the second model, DBN is trained using fresh, unlabeled data, giving it   more information\\
         && on how malicious attack patterns have changed.\\
         && The training process is easier and able to  simultaneously adopt new virus behaviors from readily \\
         && available, inexpensive, and unlabeled data.\\
         \\
         \cite{luo2019using} & ITSD merged with different & ITSD ruled out the limitations of the current line segment generating techniques.\\
         &Machine learning models& The experimental findings and mathematical analysis show that it has the potential to \\
         &&  numerous anomalies.\\
         \\
         \cite{8005871} & DBN \& PNN & Initially raw data is transformed to low dimensional data by using non-linear learning\\
         & & ability of DBN. PNN is used to classify the low dimensional data\\
         \bottomrule
    \end{tabular}
\end{table*}

The load demand forecasting is the most important task in the EV charging stations to predict the number of incoming and outgoing EVs for a given duration of time. The authors Cui et al. of \cite{cui2019machine} divided the load forecasting attacks into five categories pulse, scaling, ramping, random and smooth-curve attacks.

A comparison for the physics based and ResNet (Residual Networks) auto encoder based anomaly detection was done in \cite{mavikumbure2023physical}, and the results conclude that the ResNet based auto encoding method shows a high accuracy of 96.82\%. A novel multi head based anomaly detection method was proposed to cast out the complex and heavy NN structure \cite{li2020detecting}, and the results show that the proposed model has the highest accuracy (99.86\%) and F scores. This method overrides the manual feature extraction for multi dimensional data and also because the feature extraction mostly relies on matrix dot multiplication, The operation will be accomplished by using computational resources like graphical processing units (GPU) to accelerate it.

To overcome the problem of limited abnormal data, Luo et al. proposed imbalanced triangle synthetic data (ITSD) based on synthetic minority over-sampling technique (SMOTE) that brings a balanced effect on normal and abnormal data and this technique has the capability of merging with different machine learning algorithms for anomaly prediction. Zhao et al. introduced an intrusion detection method using a combination of deep belief network (DBN) and probabilistic neural network (PNN) \cite{8005871}. The use of DBN, with its capacity for nonlinear learning, has yielded superior performance when compared to the conventional PNN approach. The determination of the hidden layer's node count is accomplished through the utilization of the particle swarm optimization (PSO) method.

A framework for smart connected automobiles that supports automated secure continuous cloud service availability, enables an IDS against security threats, and offers services that satisfy user quality of service (QoS) and quality of experience (QoE) needs. Smart vehicles are grouped into service-specific clusters to achieve continuous service availability. To facilitate communication between service requesters and suppliers, trustworthy third-party (TTPs) entities are chosen as the cluster leaders. Additionally, a three-phase data traffic analysis, reduction, and classification technique is employed to identify the positive trusted service requests against fraudulent ones, that might occur during intrusion attacks to achieve intrusion detection. For data reduction and classification, the system uses DNN and decision tree machine learning algorithms, respectively. 
Simulations serve as a validation mechanism for the proposed framework, demonstrating its effectiveness in the context of intrusion attack detection. The suggested solution exhibits notable performance metrics, with a 99.43\% overall accuracy, a false positive rate of 0.96\%, a false negative rate of 1.53\%, and a detection rate of 99.92\%.

\subsection{Defense Mechanisms for Cyber Attacks}
During the past few decades, the evolution of automotive systems from electromechanical to digital electronic and software dependent systems has changed the dynamics of the EV industry. The article in the literature \cite{du2022review}, divides the defense strategies of CPPS  into two categories: passive defense strategies, which focus on swiftly isolating the attack affected regions, and taking the necessary countermeasures to ensure the normal operation of the CPPS. The active defense strategies are the ones which aim to completely eliminate the likelihood of any successful attacks. Numerous corresponding protection strategies have been created in order to further increase the security of CPPS and lessen the possibility of cyberattacks. 
 
The abnormality in the battery charging capacity may also be caused due to these cyber attacks. Various types of anomaly diagnosis methods are available, that are approximately classified as knowledge-based, model-based, and data-driven methods \cite{wang2021data}. For knowledge based methods, a method is proposed where the design of fault diagnosis rules is done by using the prior expert and then applied to several battery parameters that are easily measured. The diagnosis of the EVs of the same model is done by the same rules \cite{zhao2016simulation}. The model based methods use the physics based models representing the electrochemical or thermal models. The overall defense strategies are divided up into two areas in this article: prevention and mitigation strategies.

\subsubsection{Mitigation Defense Strategies}
These type of defense strategies should be able to be implemented after the occurrence of an attack. When an attack occurs, the affected area or the node must be in solitude at an early stage,  later certain measures and actions will be performed to reduce the damage caused and also for the system restoration. For the secure and dependable functioning of the system, quick detection and identification of cyber-attacks are essential. By using a defense-in-depth strategy to identify, stop, and protect the system against these threats, the attacks are prevented.

Chen-ching Lu proposed a mitigation methodology in four steps \cite{liu2011intruders}, namely, modeling of the CPPS, simulation of the physical behavior of grids, development of vulnerability index for the system, and the mitigation methods. Mitigation is performed on the information and communication technology (ICT) side of the CPS and also on the grid side. On the ICT side, it is achieved by using dynamic and other enhanced firewall structures, like adaptive rejection practices rules, or by restricting access via the firewall. Computational algorithms make a path for finding the reconfiguration of the power system network during an attack. This work doesn't include time synchronization attacks and more industrial communications protocols might be incorporated. 

\subsubsection{Prevention Defense Strategies}
For the purpose of reducing and eliminating the impacts of the threat caused by cyber attacks, preventive measures should be taken prior to the attack. Unlike the aforementioned, this prevention strategy stops the cyber attacks ahead of their occurrence, so that the effect of the attack should be nullified at a prior stage. There are several methods available in the literature that are adopted for the prevention of cyber attacks effect, but most of the existing methods are applicable to the power system transmission lines \cite{du2022review, tian2018enhanced, hasan2020game, xiang2018improved} and the cascading effects caused by the cyber attacks as explained in the above sections will be prevented by improving the resilience of the system \cite{babu2023resilient}. A data-driven methodology is presented for enhancing resilience by improvement vulnerable nodes within the system. These identified nodes can then be enhanced by implementing EVCI and outfitting them to function as DER with an integrated energy management system (EMS) \cite{reddy2023data}. The allocation of EV charging stations at different vulnerable locations is discussed in \cite{chakraborty2024planning}. Apart from the above-mentioned methods, other resilience improvement methods may be used for the prevention of cyber attacks \cite{dwivedi2023partitioning}.

Defending against cyber attacks is essential to protect our systems, data, and information from malicious attackers. The defense methods opted for various strategies that encompass distinct features, advantages, and disadvantages. The various benefits and drawbacks of the aforementioned cyber attack defense strategies \cite{du2022review} are discussed in Table \ref{defense strategies}.

\begin{table*}[]
    \centering
    \caption{Benefits and drawbacks of various cyber attack defense strategies}
    \label{defense strategies}
    \begin{tabular}{p{3.5cm}p{6cm}p{7cm}}
    \toprule
         \textbf{Category} & \textbf{Advantages} & \textbf{Disadvantages} \\
         \midrule
         Mitigation strategies & \begin{enumerate}
             \item Rapid identification of compromised nodes.
             \item Uninterrupted system operation during the attacked mode.
         \end{enumerate} & \begin{enumerate}
             \item Susceptible to attack tolerance delays during attacks.
             \item Simplifies the aggravation of system instability.
             \item Errors in isolating the unattacked nodes.
         \end{enumerate}\\
         \midrule
         Prevention strategies &\begin{enumerate}
             \item No complexities in the defender operations.
             \item Underutilization of defensive services.
         \end{enumerate} &\begin{enumerate}
             \item Discrepancy in the treatment of targeted and safeguarded nodes.
             \item Disparities in resource allocation for attacking and defending the same target.
         \end{enumerate}\\
         \bottomrule
    \end{tabular}
\end{table*}

\section{Conclusions and Future Scope}
\label{section:Conclusion}
This comprehensive review focuses on the various aspects and considerations that model the EVCI as the CPS. This article explores the EVCI within the context of CPS in the societal landscape, highlighting its significance in shaping the future of the transportation and energy sectors. Few major cyber incidents are discussed with their impacts. A typical structure of the EVCI is introduced comprising the cyber and physical layers of the system. The different parts of the EVCI discussed are EVSE, CS, EV, and CSMS, etc., Furthermore, the power flow and the communication flow are also provided along with the utility. The developments in the EVCI are also explored. The charging mechanisms are categorized as on-board and off-board charging systems. Also, the DC fast chargers for HDV and MDV are also explained. The role of communications and battery technologies in EVCI are also provided. This article categorizes the payment schemes and methods as fixed-rate, time-based, energy-based, hybrid and advanced pricing models. The switchgear technologies and the components required for the EVCI are also listed, which distinguish the traditional switchgear from the smart switchgear technologies. 

Various standards and protocols used in EVCI are discussed and classified based on the various standards organizations. These standards are used at the various stages in EVCI such as grid integration, off-board and on-board charging compatibility, types and modes of connectors and sockets used, interfacing compatibilities, supply circuits, power converters, safety issues, etc., A few examples of charging connectors and chargers such as CHAdeMO, CCS, and Guobiao standards are also specified.  The communication protocols are classified into front-end and back-end categories, and detailed explanations are provided regarding the specifications and versions of OCPP. The different concepts of cybersecurity such as objectives, impacts, requirements, and challenges are explained along with the CIA triad objective. The cybersecurity impacts are categorized into two groups: technical and economic impacts. Within the technical impacts, there is a further breakdown into cascading failures and stability operation impacts. The AAA framework is elucidated as a fundamental prerequisite in the realm of cybersecurity, integral to the establishment of robust security postures. Furthermore, the implementation of cybersecurity endeavors encounters a myriad of challenges, delineating formidable obstacles that necessitate strategic mitigation and resolution.

Vulnerabilities within the EVCI are examined using the cyber intrusion spiral framework, shedding light on specific susceptible domains. A comprehensive analysis of various potential attack vectors such as jamming, FDIA, spoofing, DoS, replay attacks, and more is provided, along with a systematic classification based on the model layers where these attacks manifest, encompassing both the cyber and physical layers. The primary objective of this paper is to elucidate the characteristics of EVCI when viewed as a CPS, while also delving into attack detection and defense strategies. This article examines the shared attributes within instances of attacks, intrusions, and anomalies, and concurrently investigates a spectrum of detection mechanisms. Subsequently, it classifies defense mechanisms into two principal strategies, specifically mitigation and prevention methods, and furnishes a comprehensive assessment of their respective merits and drawbacks.

After a comprehensive review of the relevant literature, several research gaps in the existing research landscape become apparent, necessitating further exploration of novel technologies and solutions but also delineating the potential avenues for future endeavors:
 
\begin{enumerate}
    \item Since the EVCI is at an initial stage in implementation, zero-day attacks play a major role and hence are given a high priority in order to make the EVs and the EVCI resilient by preventing malware transmission and regular operations.
    \item Periodic enhancement or establishment of novel standards and protocols is imperative to mitigate unauthorized access by potential adversaries in accordance with the evolving requirements of EVCI systems.
    \item Since the coordination of the EVCI with the grid and the power optimization methods are beyond the scope of this article, there are still certain improvements and modifications that need to be done for the resilient and economic operations of the EVCI administrator, EV user, and DSO operator.
    \item The continuous maintenance and refinement of diverse pricing methodologies and mechanisms necessitates the adoption of strategies such as ToU and dynamic pricing, which are instrumental in enabling intelligent charging solutions.
    \item The optimal allocation of time slots for EVs at public charging stations relies on a comprehensive assessment of factors including traffic patterns and charging duration. Furthermore, it is imperative for charging strategies to encompass demand-side management while concurrently offering ancillary services to improve the grid performance, encompassing tasks like voltage and frequency regulation, as well as the provisioning of active and reactive power.
    \item Advancements in the interface design for EV users, CSMS operators, and DSOs are aimed at creating automated systems that enhance user-friendliness and promote seamless system operation. These services are designed to offer various functionalities, including payment processing and slot reservation capabilities.
    \item The advancement of dynamic self-healing mechanisms is of paramount importance to ensure adaptability and an optimal architectural framework. Furthermore, there is a need for enhancements in the frameworks pertaining to defense strategies for CPS based EVCI systems.
    \item Considering the intricate stratification of CPS, it is imperative to broaden strategies for deploying multiple resources and implementing safety measures across diverse layers and systems, with the aim of augmenting the comprehensive performance of the entire system.
\end{enumerate}

\bibliographystyle{IEEEtran}
\bibliography{main}

\end{document}